\documentclass[onecolumn, prd, aps, tightenlines, preprintnumbers, showpacs, nofootinbib, superscriptaddress, notitlepage]{revtex4-1}

\pdfoutput=1

\usepackage{float}
\usepackage{amsmath}
\usepackage{color}
\usepackage{graphicx}
\usepackage[dvipsnames]{xcolor}
\usepackage{url}
\usepackage{epsfig}
\usepackage[T1]{fontenc}
\usepackage{multirow}
\usepackage{physics} 
\usepackage{booktabs} 
\usepackage{array} 
\usepackage{paralist} 
\usepackage{grffile}
\usepackage{verbatim} 
\usepackage{subfig} 
\usepackage{amsmath,amsthm,amssymb,bm,amsfonts}
\usepackage{slashed}
\usepackage[utf8]{inputenc}
\usepackage{hyperref}
\allowdisplaybreaks[1]

\usepackage{color}
\usepackage[normalem]{ulem}

\def\beg{\begin{equation}}
\def\eeg{\end{equation}}
\def\bea{\begin{eqnarray}}
\def\eea{\end{eqnarray}}

\usepackage{multirow}
\usepackage[title]{appendix}

\def\Tr{{\rm tr}}

\newcommand{\slv}{\raise.15ex\hbox{$/$}\kern-.53em\hbox{$v$}}
\newcommand{\slnbar}{\raise.15ex\hbox{$/$}\kern-.53em\hbox{$\bar{n}$}}
\newcommand{\slF}{\raise.15ex\hbox{$/$}\kern-.53em\hbox{$F$}}
\newcommand{\sllbar}{\raise.15ex\hbox{$/$}\kern-.40em\hbox{$\bar{l}$}}
\newcommand{\slh}{\raise.15ex\hbox{$/$}\kern-.40em\hbox{$h$}}
\newcommand{\slP}{\raise.15ex\hbox{$/$}\kern-.53em\hbox{$P$}}
\newcommand{\slR}{\raise.15ex\hbox{$/$}\kern-.53em\hbox{$R$}}
\newcommand{\slz}{\raise.15ex\hbox{$/$}\kern-.53em\hbox{$Z$}}
\newcommand{\slzbar}{\raise.15ex\hbox{$/$}\kern-.53em\hbox{$\bar{Z}$}}
\newcommand{\slQ}{\raise.15ex\hbox{$/$}\kern-.53em\hbox{$Q$}}
\newcommand{\slK}{\raise.15ex\hbox{$/$}\kern-.53em\hbox{$K$}}
\newcommand{\slkbar}{\raise.15ex\hbox{$/$}\kern-.53em\hbox{$\bar{k}$}}
\newcommand{\slkone}{\raise.15ex\hbox{$/$}\kern-.53em\hbox{$k_1$}}
\newcommand{\slpone}{\raise.15ex\hbox{$/$}\kern-.53em\hbox{$p_1$}}
\newcommand{\slpbarone}{\raise.15ex\hbox{$/$}\kern-.53em\hbox{$\bar{p}_1$}}
\newcommand{\slptwo}{\raise.15ex\hbox{$/$}\kern-.53em\hbox{$p_2$}}
\newcommand{\slpbartwo}{\raise.15ex\hbox{$/$}\kern-.53em\hbox{$\bar{p}_2$}}
\newcommand{\slqone}{\raise.15ex\hbox{$/$}\kern-.53em\hbox{$q_1$}}
\newcommand{\slD}{\raise.15ex\hbox{$/$}\kern-.53em\hbox{$\!D$}}
\newcommand{\slC}{\raise.15ex\hbox{$/$}\kern-.53em\hbox{$C$}}
\newcommand{\slA}{\raise.15ex\hbox{$/$}\kern-.73em\hbox{$A$}}
\newcommand{\slSigma}{\raise.15ex\hbox{$/$}\kern-.53em\hbox{$\Sigma$}}
\newcommand{\slpartial}{\raise.15ex\hbox{$/$}\kern-.53em\hbox{$\partial$}}
\newcommand{\slcalP}{\raise.15ex\hbox{$/$}\kern-.63em\hbox{$\cal P$}}
\newcommand{\sleps}{\raise.15ex\hbox{$/$}\kern-.53em\hbox{$\epsilon$}}
\newcommand{\slepsbar}{\raise.15ex\hbox{$/$}\kern-.53em\hbox{$\overline{\epsilon}$}}
\newcommand{\slepsstar}{\raise.15ex\hbox{$/$}\kern-.53em\hbox{$\epsilon$}^\star}
\newcommand{\slS}{\raise.15ex\hbox{$/$}\kern-.73em\hbox{$S$}}

\newcommand{\bb}{\mathbf}

\newcommand{\bk}{\mathbf{k}}
\newcommand{\bp}{\mathbf{p}}
\newcommand{\bq}{\mathbf{q}}
\newcommand{\bx}{\mathbf{x}}

\newcommand{\barq}{\bar{q}}

\newcommand{\mA}{\mathcal{A}}
\newcommand{\mM}{\mathcal{M}}

\newcommand{\dtwo}[1]{\frac{\dd^2 #1}{(2\pi)^2}}

\newcommand{\p}{\prime}

\begin{document}
\title{Single Inclusive Hadron Production in DIS at Small $x$: Next to Leading Order Corrections}

\author{Filip Bergabo}
\email{fbergabo@gradcenter.cuny.edu}
\affiliation{Department of Natural Sciences, Baruch College, CUNY, 17 Lexington Avenue, New York, NY 10010, USA}
\affiliation{City University of New York Graduate Center, 365 Fifth Avenue, New York, NY 10016, USA}

\author{Jamal Jalilian-Marian}
\email{jamal.jalilian-marian@baruch.cuny.edu}
\affiliation{Department of Natural Sciences, Baruch College, CUNY, 17 Lexington Avenue, New York, NY 10010, USA}
\affiliation{City University of New York Graduate Center, 365 Fifth Avenue, New York, NY 10016, USA}


\begin{abstract}
We calculate the one-loop corrections to single inclusive hadron production in Deep Inelastic Scattering (DIS) at small $x$ in the forward rapidity region using the Color Glass Condensate formalism. We show that the divergent parts of the next to leading order (NLO) corrections either cancel among each other or lead to $x$ (rapidity) evolution of the leading order (LO) dipole cross section according to the JIMWLK evolution equation and DGLAP evolution of the parton-hadron fragmentation function. The remaining finite parts constitute the NLO ($\alpha_s$)  corrections to the LO single inclusive hadron production cross section in DIS at small $x$. 
\end{abstract}

\maketitle



\section{Introduction}\label{sec:intro}

Gluon saturation~\cite{Gribov:1984tu,Mueller:1985wy} at small $x$ as encoded in the Color Glass Condensate (CGC) formalism~\cite{Iancu:2003xm,Iancu:2002xk,Jalilian-Marian:2005ccm,Weigert:2005us,Morreale:2021pnn} has been the subject of intense theoretical studies and experimental searches. Theoretical work based on leading order (LO) or leading log (LL) approximations to gluon saturation have successfully described structure functions, suppression of the single inclusive hadron transverse momentum spectrum and disappearance of the away side peak in dihadron angular correlations in high energy proton(deuteron)-gold/lead collisions at RHIC and the LHC~\cite{Kovner:2001vi,Jalilian-Marian:2004vhw,Dumitru:2005gt,Jalilian-Marian:2005qbq,Marquet:2007vb,Albacete:2010pg,Stasto:2011ru,Lappi:2012nh,Jalilian-Marian:2012wwi,Jalilian-Marian:2011tvq,Zheng:2014vka,Stasto:2018rci,Albacete:2018ruq,Mantysaari:2019hkq,Hatta:2020bgy,Jia:2019qbl,Gelis:2002fw,
Dominguez:2011wm,Metz:2011wb,Dominguez:2011br,Iancu:2013dta,Altinoluk:2015dpi,Hatta:2016dxp,Dumitru:2015gaa,Kotko:2015ura,Marquet:2016cgx,vanHameren:2016ftb,Marquet:2017xwy,Dumitru:2018kuw,Dumitru:2001jn,
Dumitru:2002qt,Mantysaari:2019csc,Salazar:2019ncp,Boussarie:2021ybe,Ayala:1995hx,Jalilian-Marian:1996mkd,
Kotko:2017oxg,Hagiwara:2017fye,Henley:2005ms,
Klein:2019qfb,Hatta:2021jcd,Kolbe:2020tlq,Gelis:2002fw,Altinoluk:2019fui,Boussarie:2019ero,Boussarie:2016ogo,Boussarie:2014lxa,Dumitru:2010ak}. Nevertheless firmly establishing gluon saturation as the QCD dynamics responsible for these experimental observations requires more precise theoretical calculations. The ongoing work on improving the accuracy of leading order CGC calculations can be broadly put into three categories; higher order in $\alpha_s$ corrections to leading order results~\cite{Fadin:1998py,Chirilli:2011km,Chirilli:2012jd,balitsky:2012bs,Balitsky:2013fea,
grabovsky:2013mba,caron-huot:2013fea,kovner:2013ona,lublinsky:2016meo,caron-huot:2016tzz,boussarie:2017dmx,beuf:2022ndu,beuf:2021srj,beuf:2021qqa,mantysaari:2021ryb,mantysaari:2022bsp,mantysaari:2022kdm,lappi:2021oag,Iancu:2020mos,roy:2019hwr,hatta:2022lzj,Iancu:2021rup,Taels:2022tza,Caucal:2021ent,Bergabo:2021woe,Bergabo:2022tcu}, sub-eikonal corrections which aim to relax the infinite energy assumption inherent to eikonal approximation~\cite{Kovchegov:2017lsr,Cougoulic:2019aja,Kovchegov:2018znm,Kovchegov:2017jxc,Kovchegov:2016zex,Kovchegov:2016weo,Kovchegov:2015pbl,Agostini:2019hkj,Agostini:2019avp,Altinoluk:2015xuy,Altinoluk:2015gia,Altinoluk:2014oxa}, and inclusion of intermediate/large $x$ dynamics into CGC in order to generalize CGC to include DGLAP evolution and collinear factorization and high $p_t$ physics~\cite{jalilian-marian:2021lhe,Jalilian-Marian:2019kaf,Jalilian-Marian:2018iui,Jalilian-Marian:2017ttv,Hentschinski:2017ayz,Hentschinski:2016wya,Gituliar:2015agu,Balitsky:2016dgz,Balitsky:2015qba}. Here we will focus on next to leading order corrections to single inclusive hadron production in Deep Inelastic Scattering (DIS) at small $x$ in the forward rapidity region~\cite{Marquet:2009ca} (virtual photon going direction) for the case when the virtual photon is longitudinal. We note that leading order results for single inclusive hadron production in DIS in the midrapidity region were obtained in~\cite{Kovchegov:2001sc}. 

The ideal environment in which to investigate gluon saturation and CGC is DIS experiments at high energy as the incoming virtual photon does not interact strongly. Single inclusive hadron production in DIS (SIDIS) at small $x$ is one of the most attractive channels for gluon saturation studies as it is not sensitive to Sudakov effects which can obscure saturation dynamics in dihadron production and angular correlations. Furthermore, it is more discriminatory than the total cross section (structure functions) so that it contains more information about the QCD dynamics of the target. While there exists leading order calculations of single inclusive hadron production in DIS at small $x$ in the CGC framework~\cite{Marquet:2009ca,Kovchegov:2001sc} it is highly desirable and in fact urgently needed to perform a next to leading order calculation which can then be used for quantitative studies of the transverse momentum spectra of produced hadrons in DIS with proton and nuclear targets at the proposed Electron Ion Collider (EIC). 

Here we calculate the next to leading order corrections to single inclusive hadron production in DIS at small $x$ in the forward rapidity region using the Color Glass Condensate formalism. To do so we use our recent results for next to leading order corrections to dihadron production~\cite{Bergabo:2022tcu} in DIS and integrate out one of the final state partons. As expected we encounter various divergences which appear when we integrate over the phase space of the final state parton. We show that UV and soft divergences cancel among each other while the collinear divergences associated with radiation of a massless parton are absorbed into the parton-hadron fragmentation function. We show that all quadrupole terms appearing in the intermediate steps of the calculation cancel among various terms and one is left with dipoles (and squared dipoles) only. The rapidity divergences arising from integrating over longitudinal phase space of the final state parton are absorbed into evolution of the dipoles describing the target dynamics and lead to JIMWLK evolution of the leading order cross section. The remaining terms are finite and constitute the $O (\alpha_s)$ corrections to leading order single inclusive hadron production in DIS at small $x$.

\section{Leading Order Cross Section}
To get the leading order single inclusive hadron production in DIS at small $x$ we start with the quark antiquark production cross section in DIS given by

\bea
\frac{\dd \sigma^{\gamma^*A \to q\bar{q} X}}{\dd^2 \bb{p}\, \dd^2 \bb{q} \, \dd y_1 \, \dd y_2} &=& \frac{ e^2 Q^2(z_1z_2)^2 N_c}{(2\pi)^7} \delta(1-z_1-z_2)\int \dd^8 \bx \left[S_{122^\prime 1^\prime} - S_{12} - S_{1^\prime 2^\prime} + 1\right] \nonumber \\
&& e^{i\bb{p}\cdot(\bb{x}_1^\prime - \bb{x}_1)} e^{i\bb{q}\cdot(\bb{x}_2^\prime - \bb{x}_2)} 
\bigg[4z_1z_2K_0(|\bb{x}_{12}|Q_1)K_0(|\bb{x}_{1^\prime 2^\prime}|Q_1) + \nonumber \\
&&  (z_1^2 + z_2^2) \,
\frac{ \bb{x}_{12}\cdot \bb{x}_{1^\prime 2^\prime}}{|\bb{x}_{12}| |\bb{x}_{1^\prime 2^\prime}|} \, 
K_1(|\bb{x}_{12}|Q_1)K_1(|\bb{x}_{1^\prime 2^\prime}|Q_1) 
\bigg] .
\eea
\noindent where ($\bp , y_1$) and ($\bq , y_2$) are the transverse momentum and rapidity of the produced 
quark and antiquark, respectively, and $Q^2$ is the virtuality of the incoming photon. We have made the following definitions and short hand notations,

\begin{align}
Q_i = Q\sqrt{z_i(1-z_i)}, \,\,\,\,\,\, \bx_{ij} = \bx_i - \bx_j,\,\,\,\,\,\, \dd^8 \bx = \dd^2 \bx_1 \, \dd^2 \bx_2\, \dd^2 \bx_{1^\p} \, \dd^2 \bx_{2^\p}.
\end{align}

\noindent We have also defined $z_1 \equiv \frac{p^+}{l^+}\,\, , \,\, z_2 \equiv \frac{q^+}{l^+}$ as the momentum fractions carried by the final state quark and antiquark relative to the photon's longitudinal momentum $l^+$. In terms of these momentum fractions the rapidity is related via $\dd y_i = \frac{\dd z_i}{z_i}$. All the dynamics of the strong interactions and gluon saturation are contained in the dipoles $S_{ij}$ and quadrupoles $S_{ijkl}$, normalized correlation functions of two and four Wilson lines

\begin{align}
S_{ij} = \frac{1}{N_c} \Tr\left\langle V_i V_j^\dag \right\rangle, \,\,\,\,\,\,\,\,\, S_{ijkl} = \frac{1}{N_c}\Tr\left\langle V_i V_j^\dag V_k V_l^\dag\right\rangle, \label{dipquad}
\end{align}

\noindent where the index $i$ refers to the transverse coordinate $\bb{x}_i$ and the following notation is used for Wilson lines, 

\begin{align}
V_i &= \hat{P}\exp\left( ig \int \dd x^+ A^-(x^+,\bb{x}_i)\right).
\end{align}

\noindent The Wilson lines efficiently resum the multiple scatterings of the quark and antiquark from the target hadron or nucleus. The angle brackets in Eq. \ref{dipquad} signify color averaging~\footnote{Throughout the paper we assume that these dipoles and quadrupoles are real, nevertheless both can have imaginary parts which however do not contribute here.}. It is important to keep in mind that as this is a classical result the cross section has no non-trivial $x$ (or rapidity/energy) dependence. It is also easy to check that if one integrates over the phase space of the quark and antiquark one recovers the standard expressions for the virtual photon-target total cross section at small $x$. 

 Integrating over the quark's momentum then sets  $z_1 = 1-z_2$ and $\bb{x}^\prime_1 = \bb{x}_1$ and gives 

\bea
\frac{\dd \sigma^{\gamma^*A \to \bar{q} X}}{\dd^2 \bb{q} \, \dd y_2} &=& 
\frac{ e^2 Q^2 z_2^2 (1-z_2) N_c}{(2\pi)^5} 
\int \dd^6 \bx \left[S_{2 2^\prime} - S_{12} - S_{1 2^\prime} + 1\right] \nonumber \\
&& e^{i\bb{q}\cdot(\bb{x}_2^\prime - \bb{x}_2)} 
\bigg[4z_2 (1 - z_2) K_0(|\bb{x}_{12}|Q_2)K_0(|\bb{x}_{1 2^\prime}|Q_2) + \nonumber \\
&&  \left[z_2^2 + (1 - z_2)^2\right] \,
\frac{ \bb{x}_{12}\cdot \bb{x}_{1 2^\prime}}{|\bb{x}_{12}| |\bb{x}_{1 2^\prime}|} \, 
K_1(|\bb{x}_{12}|Q_2)K_1(|\bb{x}_{1 2^\prime}|Q_2) \label{LOdsig}
\bigg] 
\eea
where the first (second) term inside the big square bracket corresponds to contribution of 
longitudinal (transverse) photons. To get the full single inclusive production cross section 
one must also consider the case when one integrates out the antiquark. It can however be 
shown that the two results are identical so that we will only integrate out the quark 
and multiple our final results by a factor of $2$. This can also be shown to be true when 
we calculate the next to leading order corrections. Therefore we will consider only the case 
when the quark is integrated out. Furthermore and as before we will consider only the case 
of longitudinal photons in this paper.

\section{One-loop corrections}

\begin{figure}[H]
\centering
\includegraphics[width=50mm]{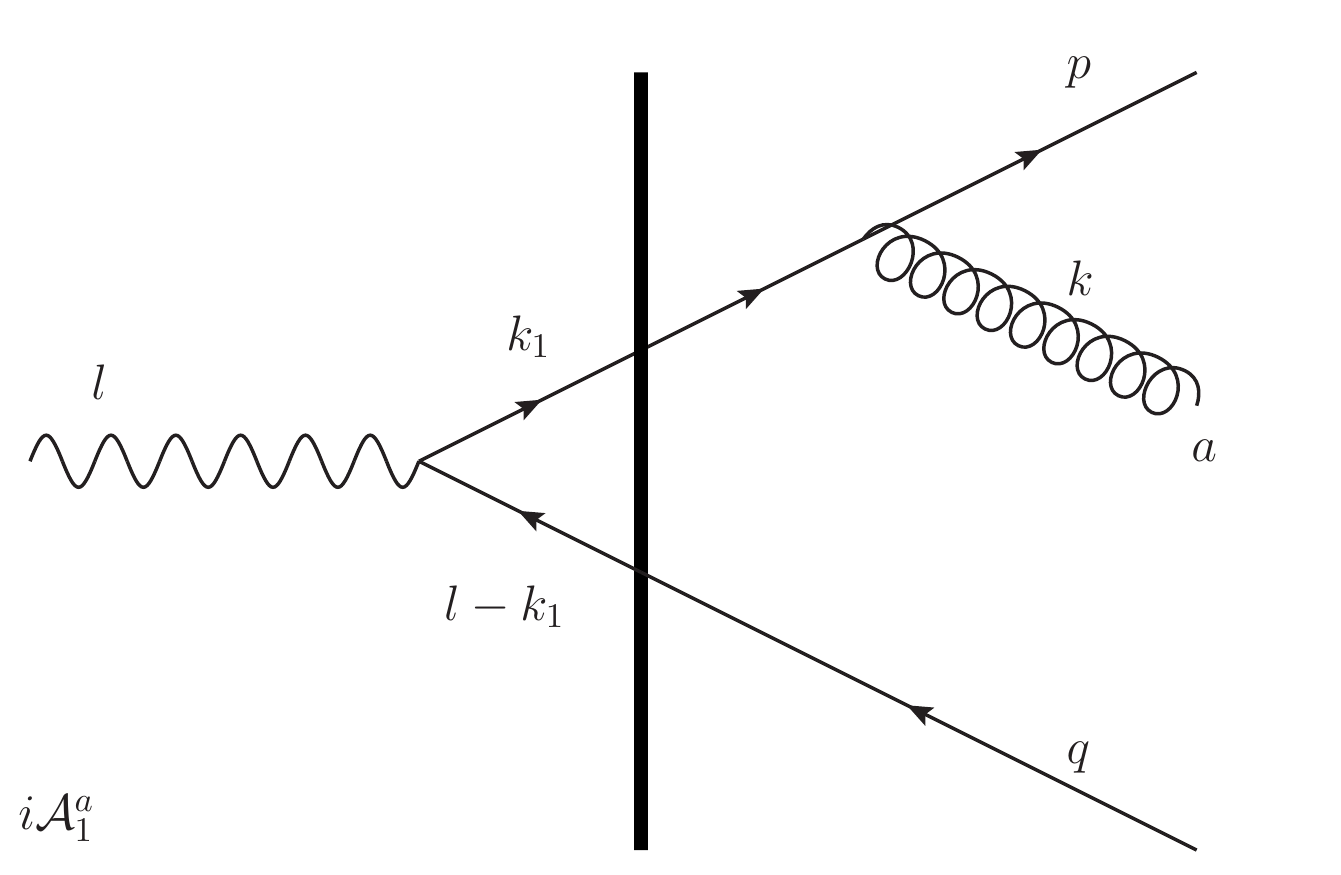}\includegraphics[width=50mm]{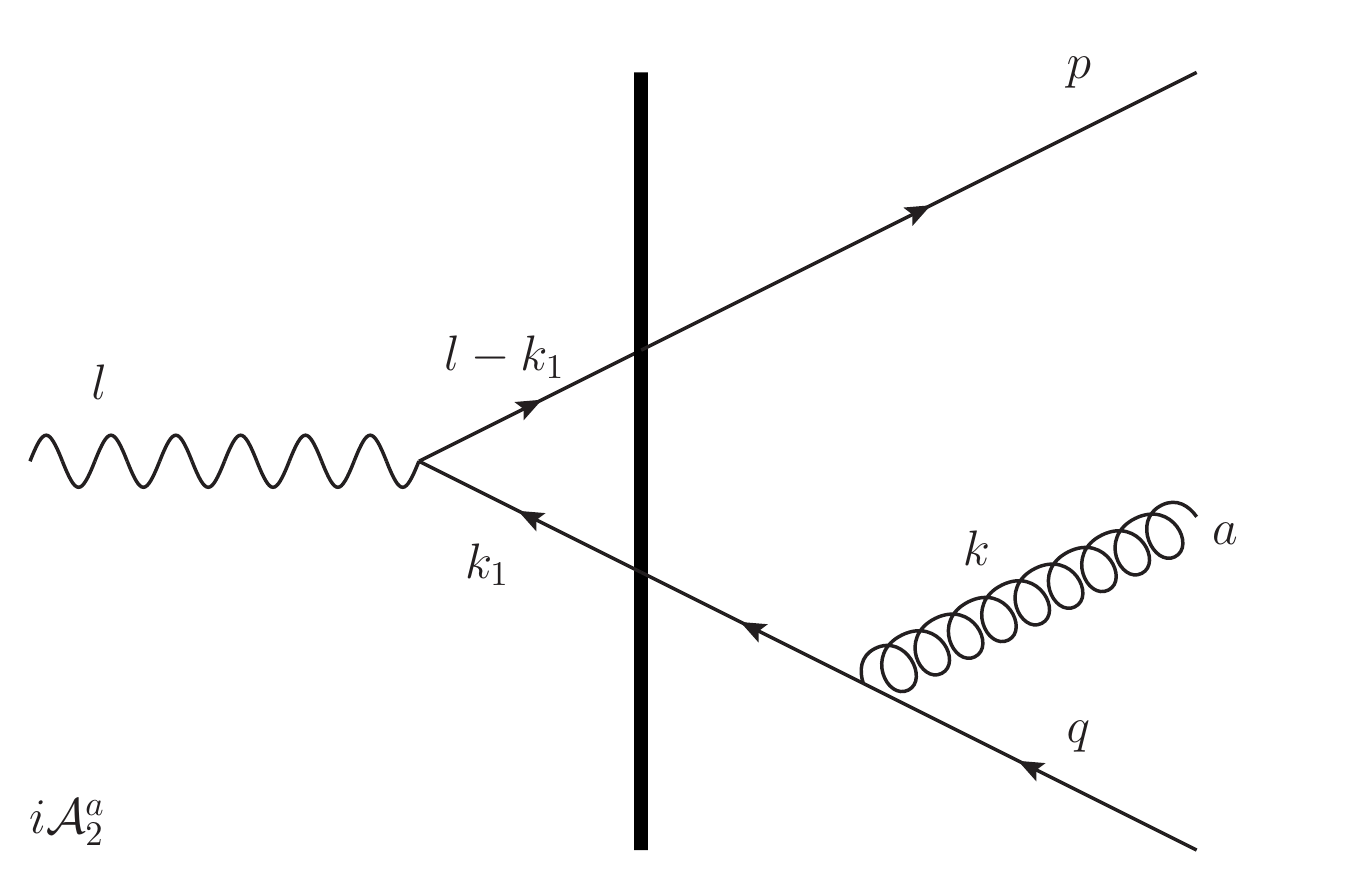}\\
 \includegraphics[width=50mm]{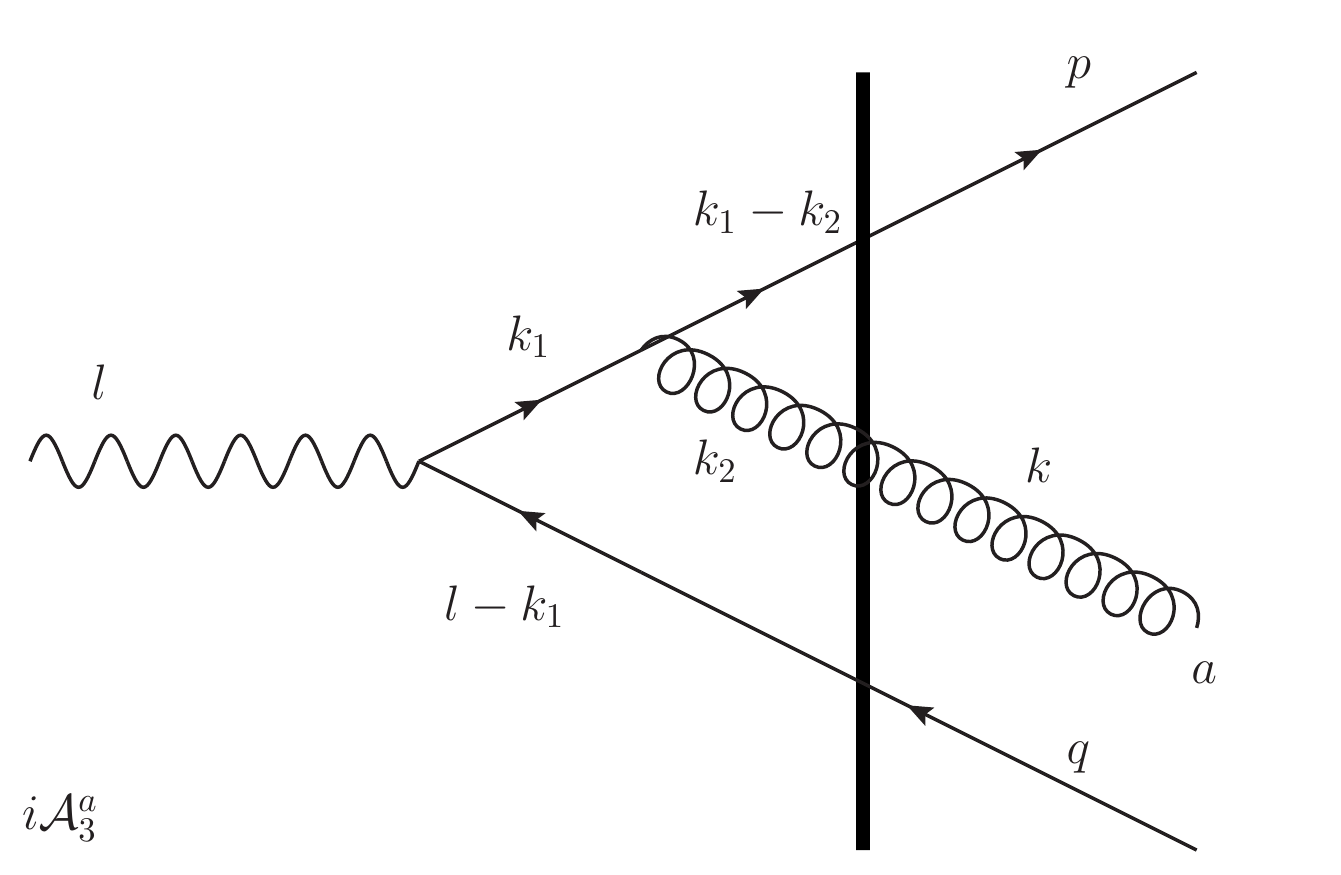} \includegraphics[width=50mm]{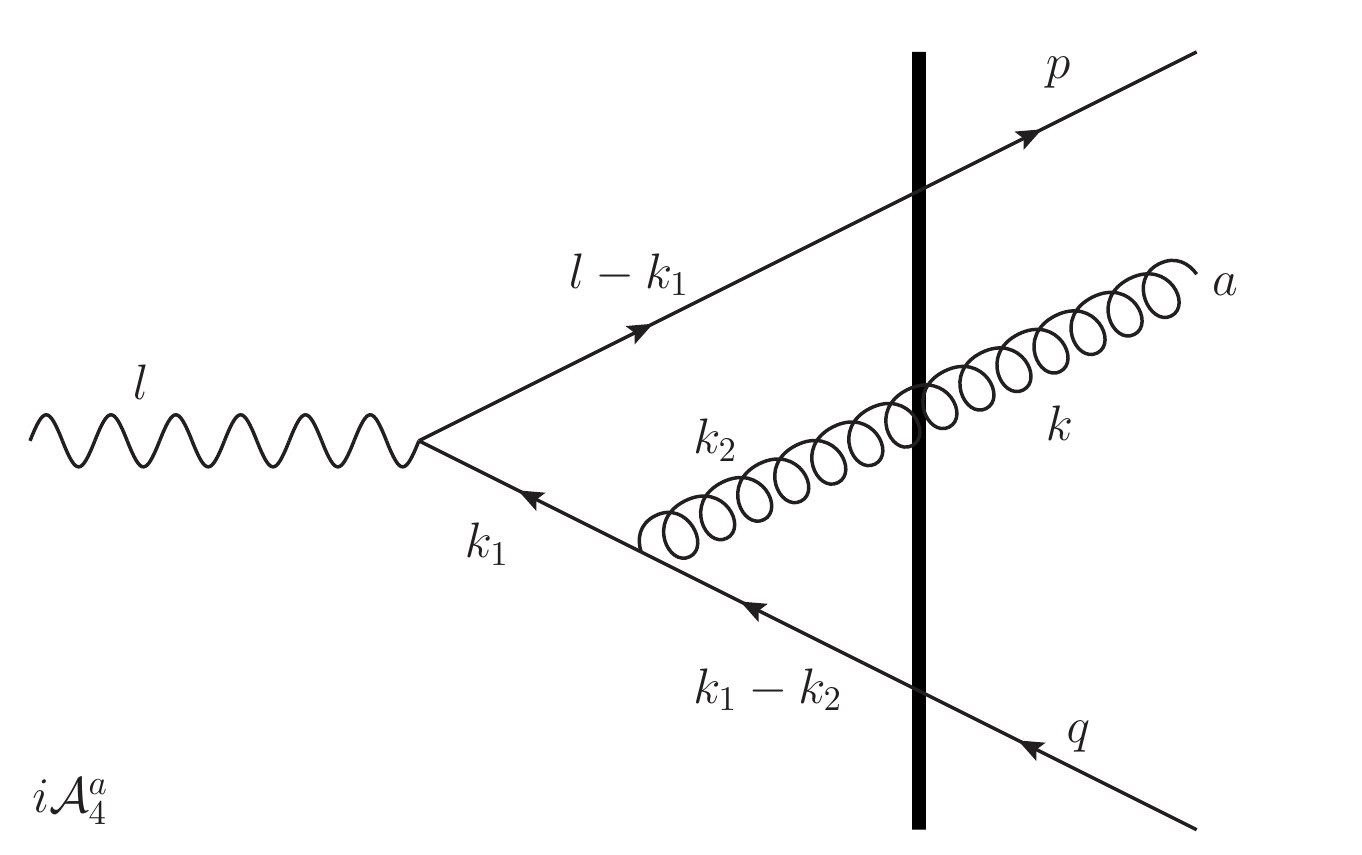}
\caption{The real corrections $i\mathcal{A}_1^a, ..., i\mathcal{A}_4^a$. The arrows on Fermion lines indicate Fermion number flow, all momenta flow to the right. The thick solid line indicates interaction with the target.}\label{fig:realdiags}
\end{figure}

\begin{figure}[H]
\centering
\includegraphics[width=52mm]{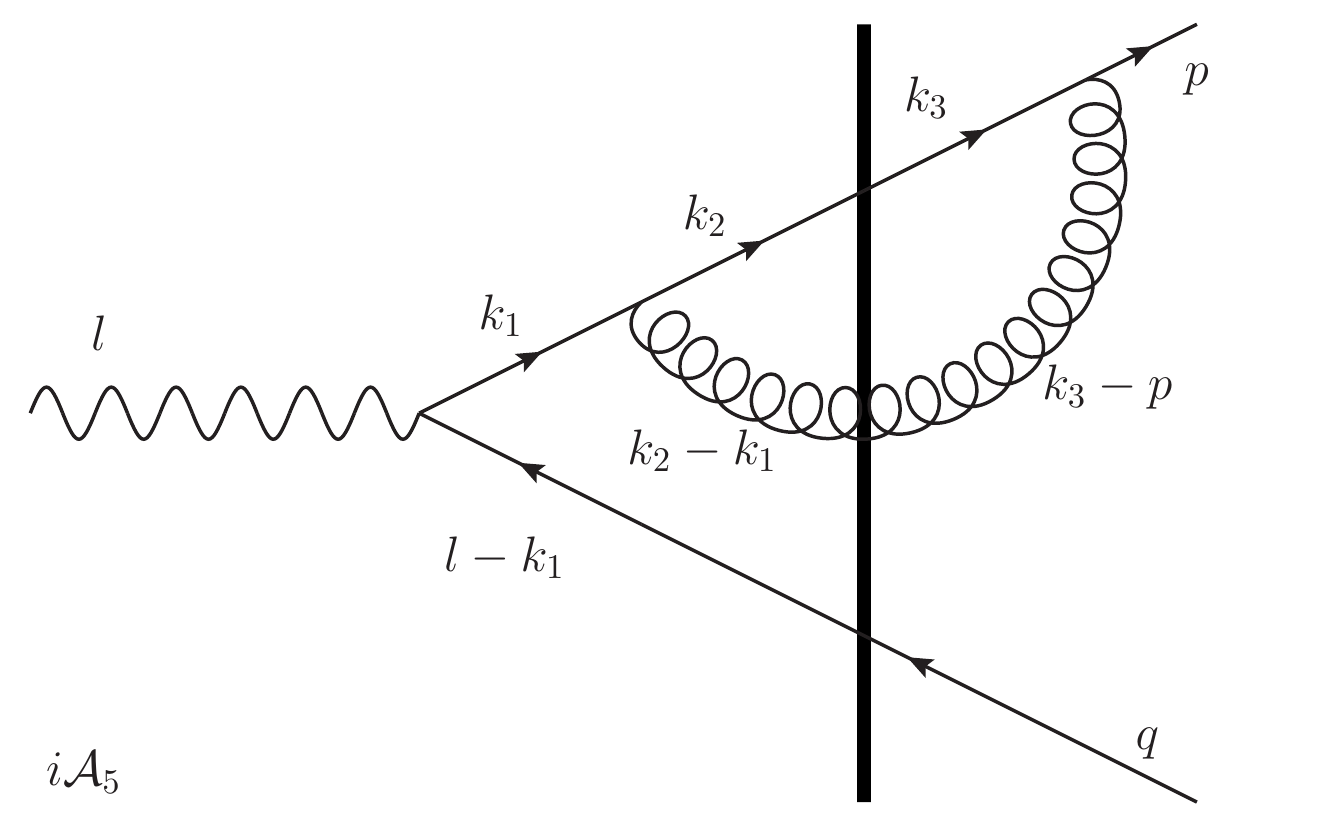}\includegraphics[width=52mm]{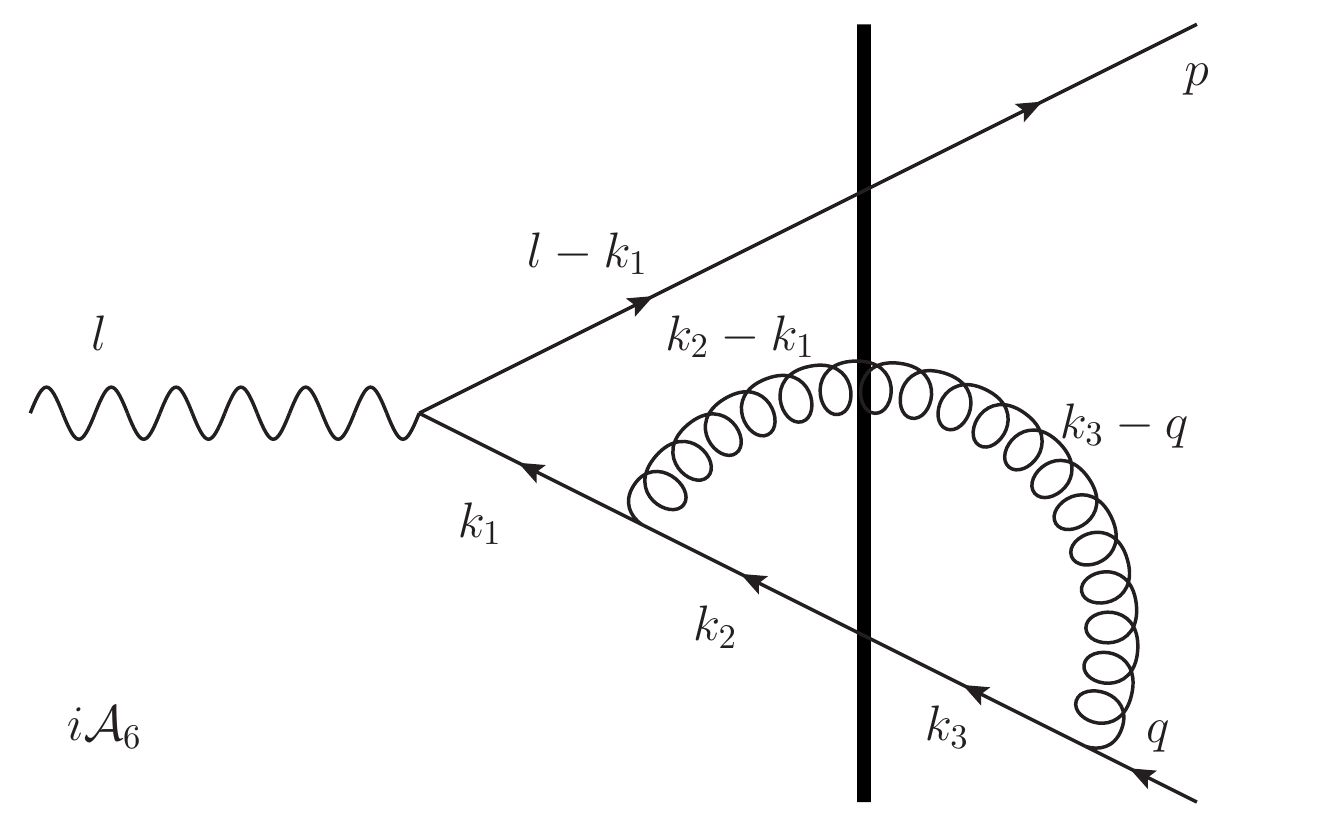} \includegraphics[width=52mm]{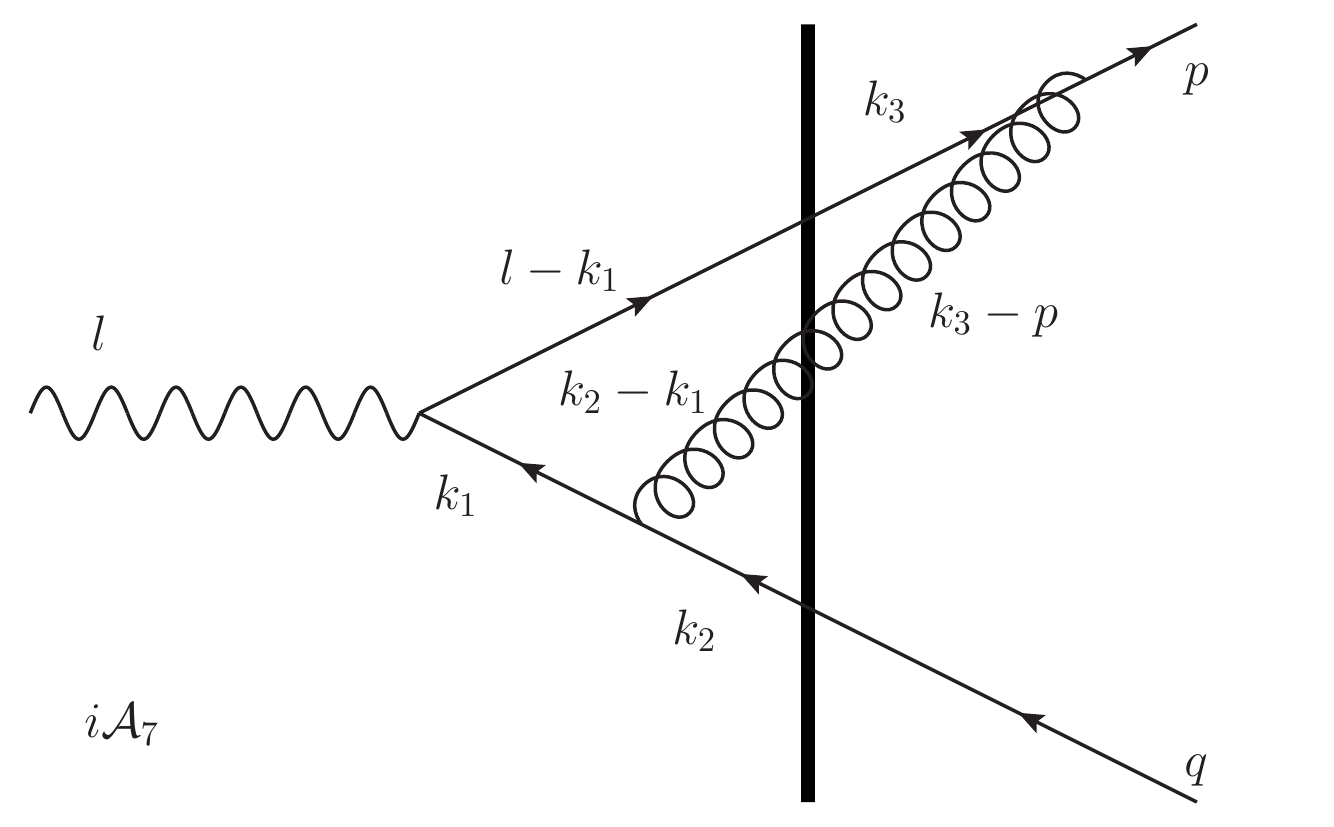} \\
\includegraphics[width=52mm]{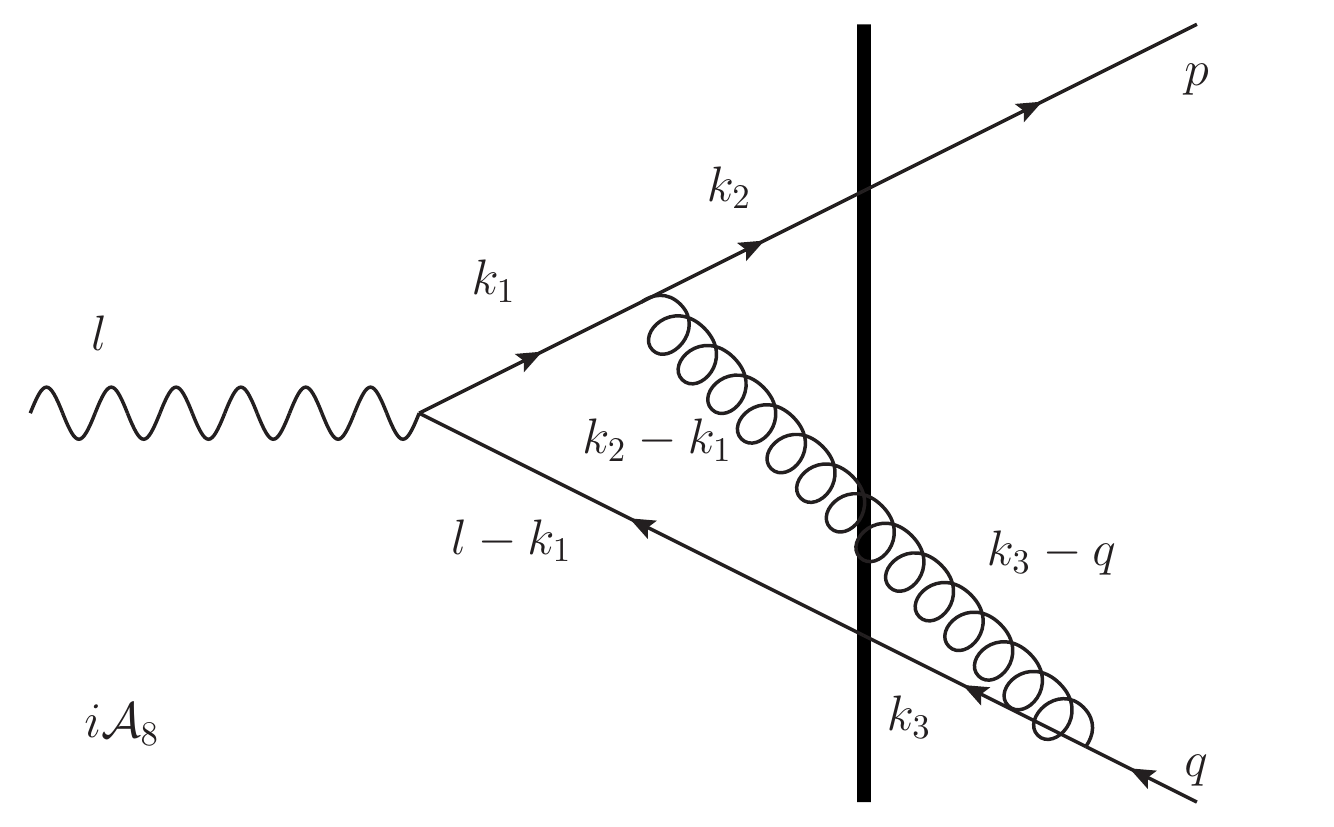} \includegraphics[width=52mm]{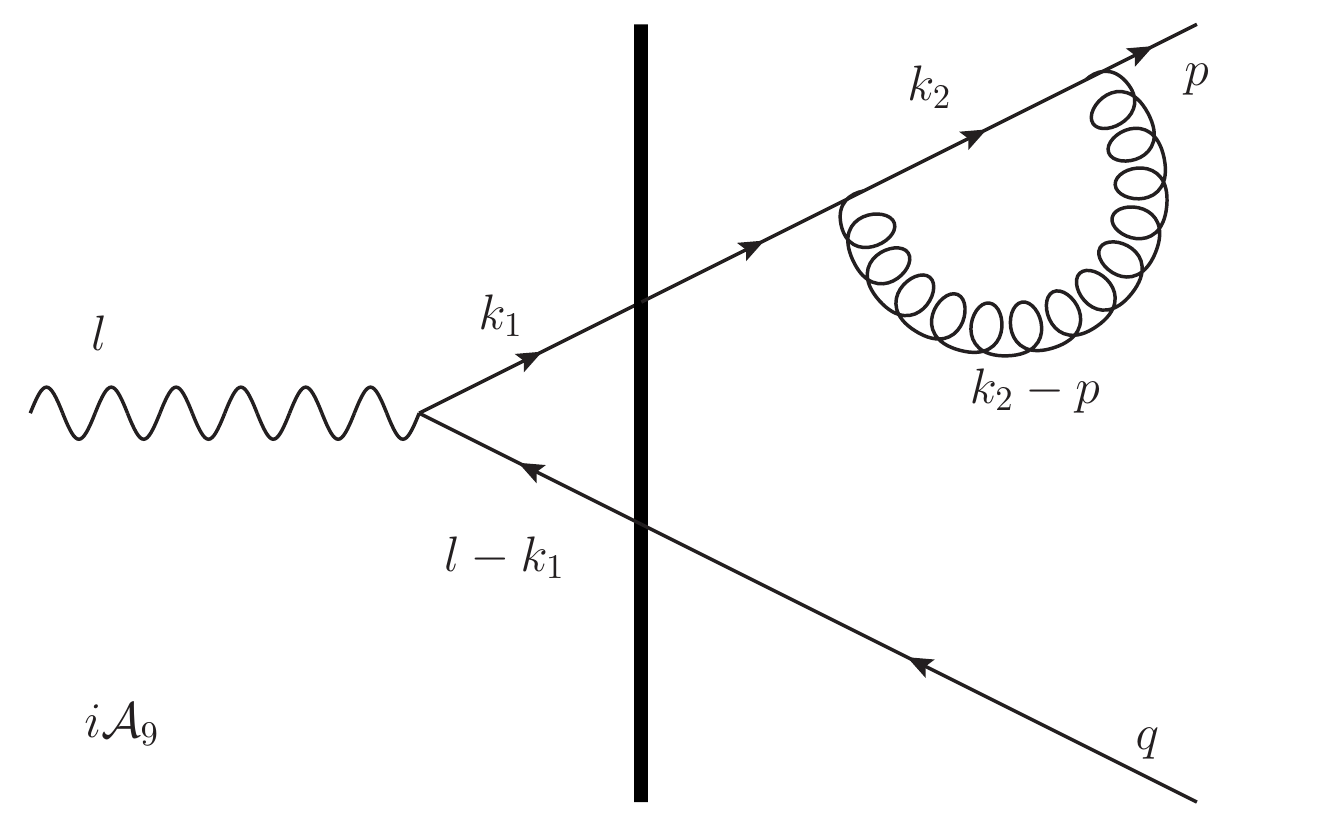}\includegraphics[width=52mm]{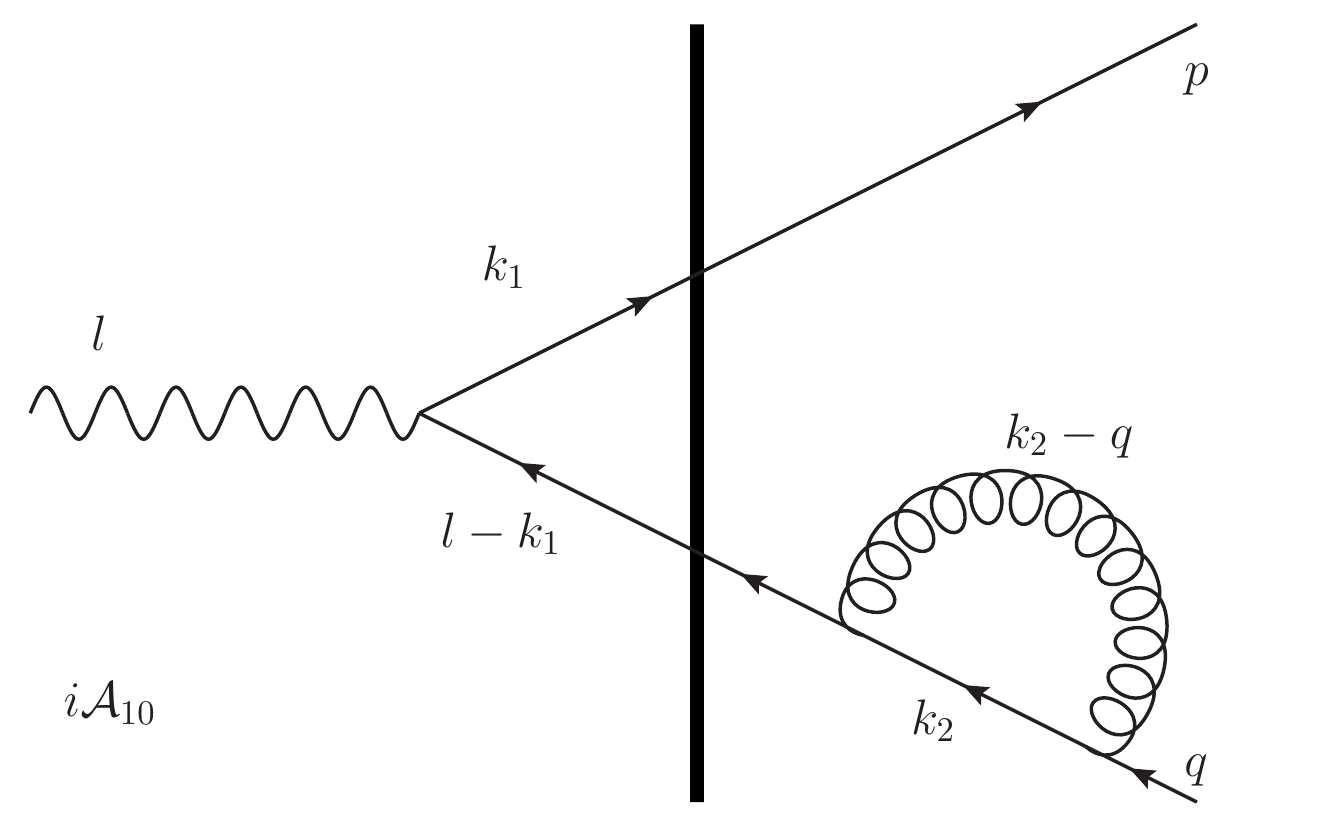}\\ 
\includegraphics[width=52mm]{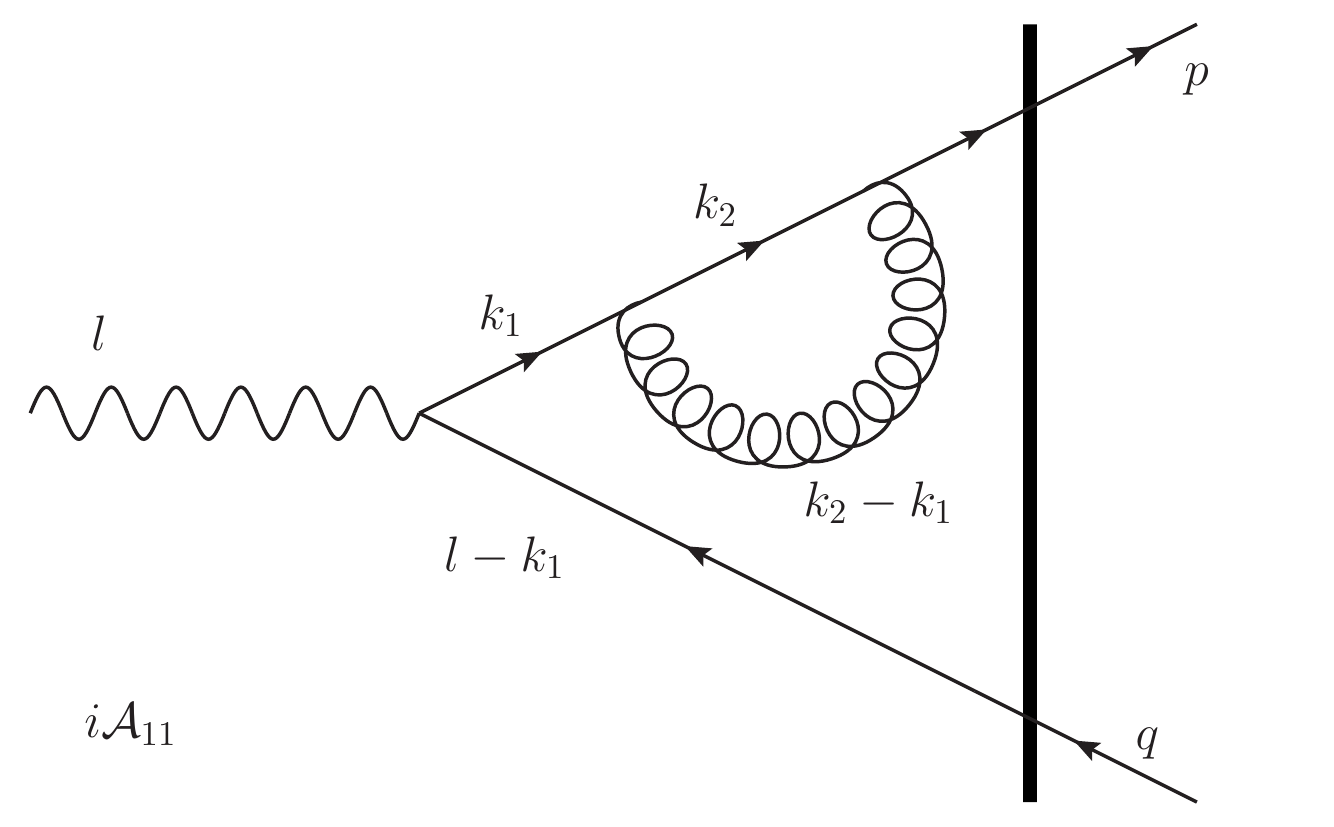}\includegraphics[width=52mm]{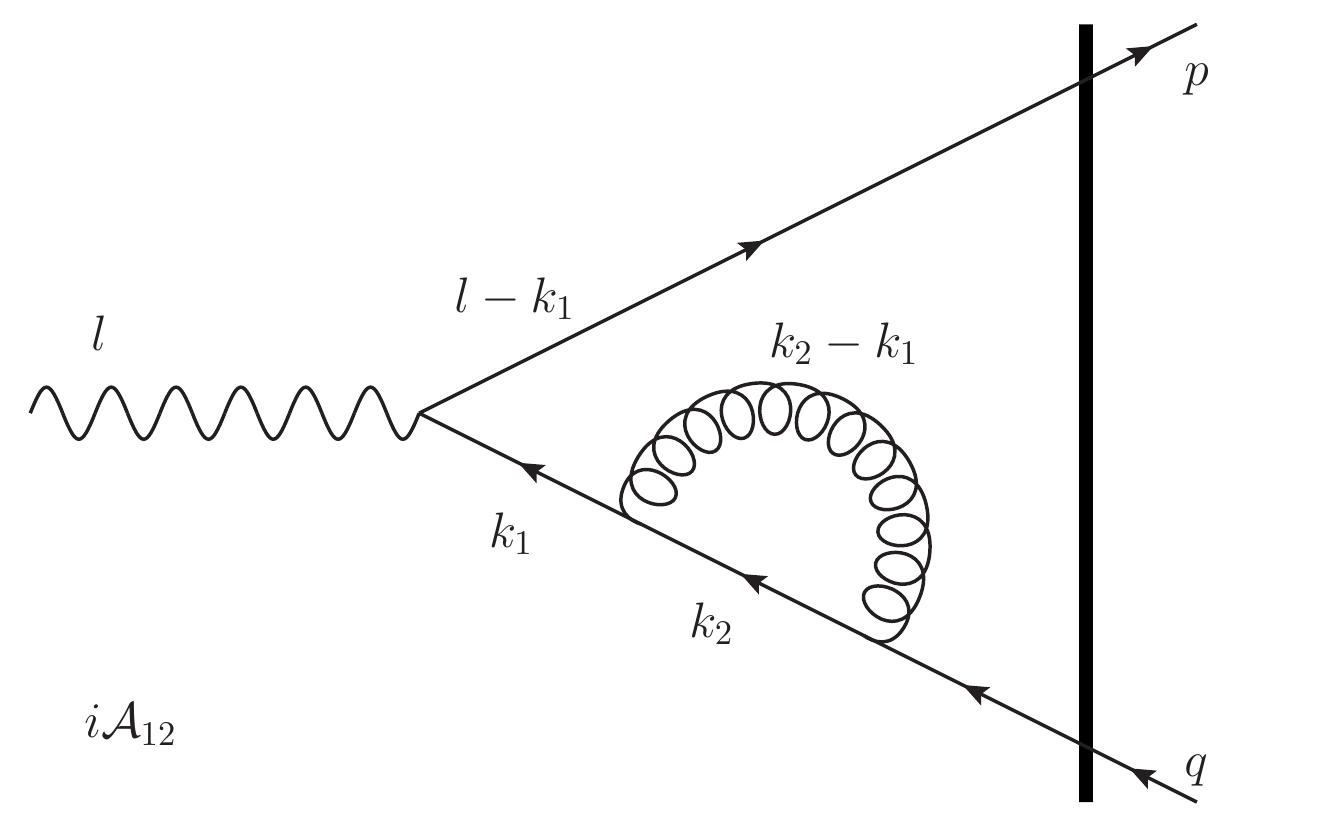} \includegraphics[width=52mm]{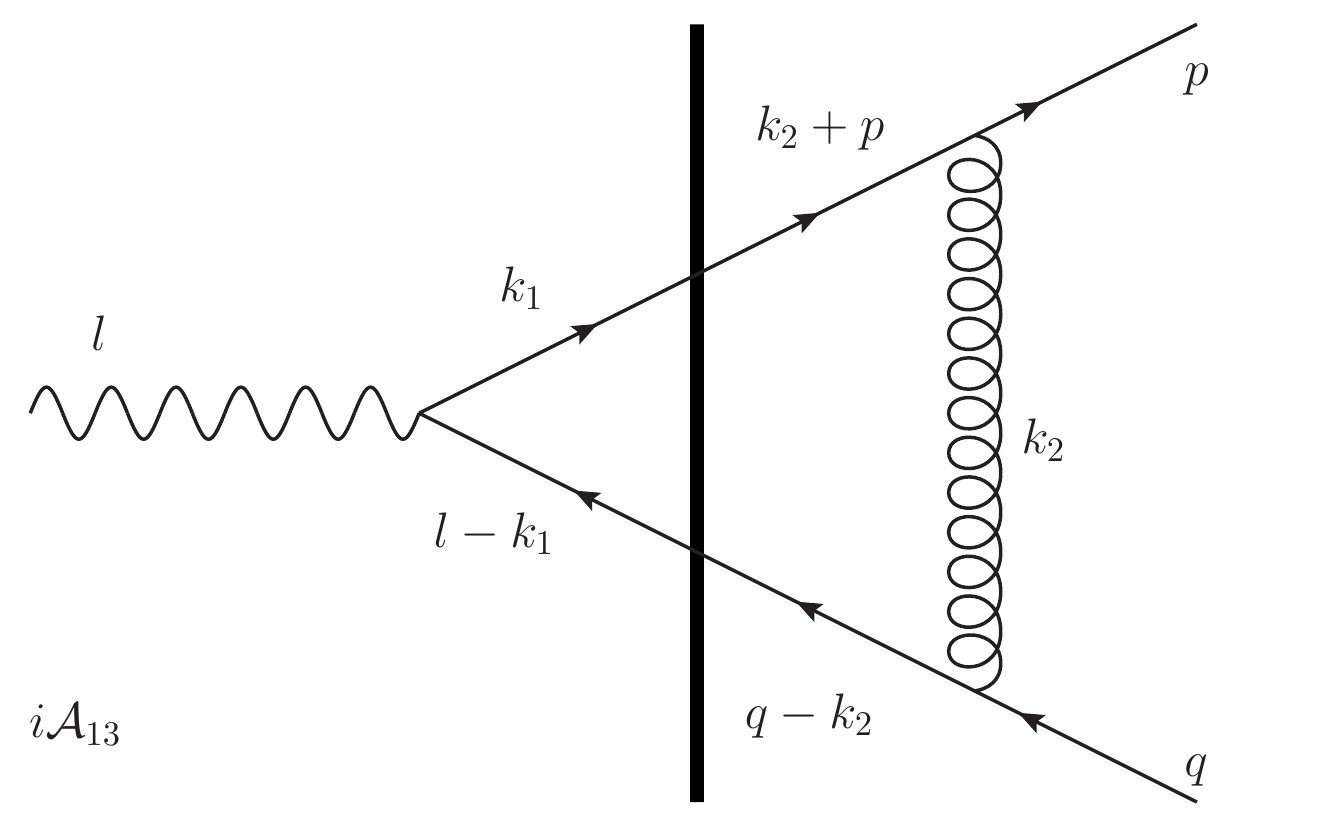}\\
\includegraphics[width=52mm]{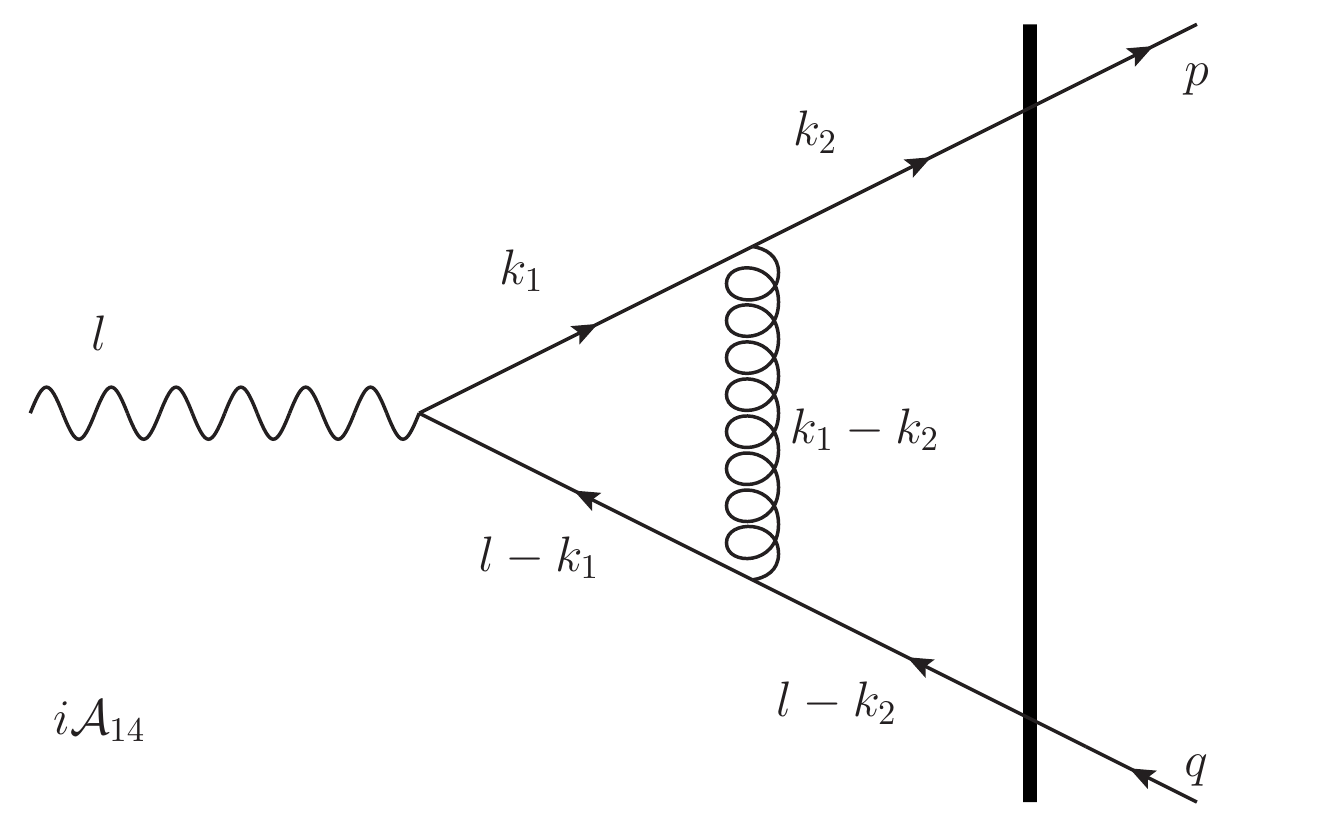}
\caption{The ten virtual NLO diagrams $i\mathcal{A}_5, ..., i\mathcal{A}_{14}$. All momenta flow to the right, \textit{except} for gluon momenta.}\label{virtualdiags}
\end{figure}

To calculate the next to leading order corrections to single inclusive hadron production 
we start with our next to leading order results for dihadron production computed in \cite{Bergabo:2022tcu}. The real corrections labeled $\dd \sigma_{i\times j}$ come from squaring the diagrams in Fig. \ref{fig:realdiags} (these were first calculated in~\cite{ayala:2016lhd,ayala:2017rmh}) and the virtual corrections $\dd \sigma_i$ from multiplying the diagrams in Fig. \ref{virtualdiags} with the leading order amplitude. These must then be multiplied by their corresponding phase space differentials $\dd\Phi^{(n)}$. The explicit details are shown in Eq. \ref{breakdown3} where we have also defined $i\mM_i^a$ via $i\mA_i^a = 2\pi\delta(l^+-p^+-q^+-k^+)i\mM_i^a$ and $i\mM_i$ via $i\mA_i = 2\pi\delta(l^+-p^+-q^+)i\mM_i$. We take the flux factor $\mathcal{F}$ to be $2l^+$. This gives the NLO corrections to quark antiquark production, and so we then integrate out the quark (i.e.: perform the integral over $\bp$ and $y_1$ in each expression) as before to obtain the single inclusive results in Eq. \ref{2x2} - \ref{142}. We note that obtaining the complete result for single inclusive hadron production starting from our original expressions for one loop corrections to dihadron production would require going back and integrating out any two of the three partons in the final state (in real corrections). As our main goal here is to demonstrate factorization of the cross section and cancellation/absorption of all divergences it is enough to focus on the case when the radiated gluon was first integrated out to get the NLO corrections to dihadron production.

\begin{align}
\dd \sigma^L_{\text{NLO}} =&\sum_{i,j=1}^4 \dd \sigma^L_{i\times j} +2\,\text{Re} \sum_{i = 5}^{14} \dd \sigma_{i}^L, \nonumber\\
\dd \sigma^{ L}_{i\times j} =& \frac{1}{\mathcal{F}} \int_{z,\bk} \left[ (i\mM_i^a)(i\mM_j^a)^{* L}\right] \dd \Phi^{(3)}, \,\,\,\,\,\, \dd \sigma^{L}_i = \frac{1}{\mathcal{F}} \left[ (i\mM_i)(i\mM)^{*,L}\right]\dd \Phi^{(2)},\nonumber \\
\dd \Phi^{(3)} =& 2l^+ \frac{ \dd^2 \bp \, \dd^2 \bq\, \dd^2 \bk \, \dd y_1 \, \dd y_2 \, \dd z}{(2\pi)^8 (4l^+)^2z} \delta(1-z_1-z_2-z), \,\,\,\,\,\,\, \dd \Phi^{(2)} = 2l^+ \frac{ \dd^2 \bp \, \dd^2 \bq\, \dd y_1 \, \dd y_2}{2(2\pi)^5 (2l^+)^2} \delta(1-z_1-z_2). \label{breakdown3}
\end{align}

\begin{align}
\frac{\dd \sigma^{L}_{2\times 2}}{\dd^2 \bq\, \dd y_2} = & \frac{2 e^2 g^2  Q^2 N_c^2 
}{(2\pi)^{8}  z_2}
\int_0^{1-z_2} \frac{\dd z}{z}(1 - z_2 - z)^2  (z + z_2)^2 
\left[z_2^2 + (z + z_2)^2\right]\int \dd^{8} \bx \, K_0(|\bx_{1 2}|Q_1) 
K_0(|\bx_{1 2^\p}|Q_1)  \Delta^{(3)}_{2 2^\p}  \nonumber \\
& [S_{2 2^\p} - S_{12} - S_{1 2^\prime} + 1] 
e^{i\bq\cdot(\bx_2^\p-\bx_2)} 
e^{i\frac{z}{z_2}\bq\cdot(\bx^\p_2-\bx_2)}\label{2x2} \\
\nonumber \\
\frac{\dd \sigma^{L}_{1\times 2}}{\dd^2 \bq\, \dd y_2} = & \frac{-2e^2 g^2  Q^2 N_c^2 
z_2}{(2\pi)^{8} }
\int_0^{1-z_2} \frac{\dd z}{z}(1-z_2-z) (z_2 + z) \left[z_2 (1-z_2)+ (1 - z_2 - z) (z_2 + z)\right]\int \dd^{8} \bx \,K_0(|\bx_{1^\p 2^\p}|Q_2)\nonumber \\ 
&K_0(|\bx_{1 2}| Q_1)\Delta^{(3)}_{1 2}
[S_{12}S_{1^\p 2^\p}-S_{12}-S_{1^\p 2^\p}+1] 
e^{i\bq\cdot(\bx_2^\p-\bx_2)} 
e^{-i\frac{z}{z_2}\bq\cdot(\bx_2 - \bx_3)}\\
\nonumber \\
\frac{\dd \sigma^{L}_{3\times 3}}{\dd^2 \bq\, \dd y_2 }=& \frac{2e^2 g^2 Q^2 N_c^2 
z_2^3 } {(2\pi)^{8}}
\int_0^{1-z_2} \frac{\dd z}{z}  \left[(1 - z_2 - z)^2 + (1 - z_2)^2\right]\int \dd^{8}\bx\, K_0(Q X) K_0(Q X_1^\p) 
\frac{1}{\bx_{31}^2}
\nonumber \\
&[S_{2 2^\p}  - S_{13}S_{23} - S_{1 3} S_{2^\p 3}+1] 
e^{i\bq\cdot(\bx_2^\p-\bx_2)} \\
\nonumber \\
\frac{\dd \sigma^{L}_{4\times 4}}{\dd^2 \bq\, \dd y_2} =& \frac{2e^2 g^2 Q^2 N_c^2 
z_2}{(2\pi)^{8}}
\int_0^{1-z_2} \frac{\dd z}{z} (1 - z_2 - z)^2 
\left[z_2^2 + (z + z_2)^2\right] \int \dd^{8}\bx \,K_0(Q X) K_0(Q X_1^\p) 
\Delta^{(3)}_{2 2^\p}
\nonumber \\
&[S_{2 2^\p} - S_{1 3}S_{2 3} - S_{1 3} S_{2^\p 3}+1] 
e^{i\bq\cdot(\bx_2^\p-\bx_2)}
\\
\nonumber \\
\frac{\dd \sigma^{L}_{3\times 4}}{\dd^2 \bq\, \dd y_2 } =& \frac{-2e^2 g^2  Q^2N_c^2 
z_2^2 }{(2\pi)^{8}}
\int_0^{1-z_2} \frac{\dd z}{z}(1 - z_2 - z) 
\left[z_2 (1-z_2) + (1 - z_2 - z) (z_2 + z)\right] \int \dd^{8}\bx 
\, K_0(Q X) K_0(Q X_1^\p) \Delta^{(3)}_{1 2^\p} \nonumber \\
&[S_{2 2^\p} - S_{1 3}S_{2 3} - S_{1 3} S_{2^\p 3}+1] 
e^{i\bq\cdot(\bx_2^\p-\bx_2)}
\\
\nonumber \\
\frac{\dd \sigma^{L}_{2\times 3}}{\dd^2 \bq\, \dd y_2} =& \frac{2e^2 g^2 Q^2 N_c^2  
z_2 }{(2\pi)^{8}}
\int_0^{1-z_2} \frac{\dd z}{z} (1 - z_2 -z) (z_2 + z) \left[(1 - z_2 - z) (z_2 + z)+ z_2 (1 - z_2)\right]\int \dd^{8}\bx 
K_0(|\bx_{1 2}|Q_1) K_0(Q X_1^\p) \Delta^{(3)}_{2 1} \nonumber \\
&[S_{2 3} S_{2^\p 3} - S_{1 3} S_{2^\p 3} - S_{12} + 1] 
e^{i\bq\cdot(\bx_2^\p-\bx_2)}
e^{i\frac{z}{z_2}\bq\cdot(\bx_3-\bx_2)}
\\ 
\nonumber \\
\frac{\dd \sigma^{L}_{2\times 4}}{\dd^2 \bq\, \dd y_2}=& \frac{-2e^2 g^2 Q^2 N_c^2  
}{(2\pi)^{8}}
\int_0^{1-z_2} \frac{\dd z}{z} (1 - z_2 - z)^2 (z_2 + z)\left[z_2^2 + (z_2 + z)^2\right]
\int \dd^{8}\bx  K_0(|\bx_{1 2}|Q_1) K_0(Q X_1^\p) 
\Delta^{(3)}_{22^\p}\nonumber \\
&[S_{2 3} S_{2^\p 3} - S_{1 3} S_{2^\p 3} - S_{1 2} + 1]  
e^{i\bq\cdot(\bx_2^\p-\bx_2)}
e^{i\frac{z}{z_2}\bq\cdot(\bx_3-\bx_2)}
\end{align}

\begin{align}
\frac{\dd \sigma^{L}_6}{\dd^2 \bq \, \dd y_2 } = &\frac{2e^2g^2Q^2N_c^2 z_2 (1 - z_2)^2 }{(2\pi)^{8}} 
\int_0^{z_2} \frac{\dd z}{z}\int  \dd^{8} \bx [S_{32^\p}S_{23} - S_{13}S_{23} - S_{1 2^\p}+1]
\left[z_2^2 + (z_2 - z)^2\right] \nonumber \\
&\frac{K_0(Q X_6)K_0(|\bx_{1 2^\p}|Q_2)}{\bx_{32}^2}
e^{i\bq\cdot(\bx_2^\p-\bx_2)}
e^{- i\frac{z}{z_2}\bq\cdot(\bx_3-\bx_2)}
\\
\frac{\dd \sigma^{L}_{8}}{\dd^2 \bq \, \dd y_2} = & \frac{-2e^2g^2 Q^2 N_c^2 z_2 (1 - z_2)}
{(2\pi)^{8}} \int_0^{z_2} \frac{\dd z \, (z_2 -z )}{z}
\int  \dd^{8}\bx\,  [S_{3 2^\p}S_{23} - S_{13}S_{23} - S_{1 2^\p}+1] 
\left[z_2 (1 - z_2) + (z_2 - z)(1 - z_2 + z)\right]\nonumber\\
&K_0(QX_6)K_0(|\bx_{1 2^\p}|Q_2) \Delta^{(3)}_{12} 
e^{i\bq\cdot(\bx_2^\p-\bx_2)}
e^{- i\frac{z}{z_2}\bq\cdot(\bx_3-\bx_2)}
\\
\frac{\dd \sigma_{10}^{L}}{ \dd^2 \bq\, \dd y_2} = &
\frac{-e^2 g^2 Q^2 N_c^2 z_2 (1 - z_2)^2}
{(2\pi)^6} \int \dd^6 \bx \big[S_{2 2^\p} - S_{12} - S_{1 2^\p} + 1\big] 
K_0(|\bx_{1 2}|Q_2) K_0(|\bx_{1 2^\p}|Q_2) 
e^{i\bq\cdot(\bx_2^\p-\bx_2)} \nonumber \\
&\times \int_0^{z_2} \frac{\dd z}{z}\, 
\left[z_2^2+(z_2-z)^2\right]
\int \dtwo{\bk} \frac{1}{\left(\bk-\frac{z}{z_2}\bq\right)^2}\label{10}
\\
\frac{\dd \sigma_{11}^{L}}{\dd^2 \bq\, \dd y_2} =& 
\frac{-2e^2g^2Q^2N_c^2 z_2^3}{(2\pi)^5} 
\int \dd^6 \bx\big[S_{2 2^\p}-S_{12}-S_{1 2^\p}+1\big] K_0(|\bx_{1 2^\p}|Q_2) 
e^{i\bq\cdot(\bx_2^\p-\bx_2)}
\nonumber\\
&\int_0^{1 - z_2} \frac{\dd z}{z} [(1 - z_2)^2 + (1 - z_2 - z)^2] 
\int \dtwo{\bk_2} \int \dtwo{\bk_1}  \nonumber\\
&\frac{e^{i\bk_1\cdot(\bx_1-\bx_2)}}{\big[\bk_1^2+Q_2^2\big]
\left[ \left(\bk_2-\frac{z}{1 - z_2}\bk_1\right)^2
+\frac{z(1 - z_2 - z)}{(1 - z_2)^2 z_2}\bk_1^2 +\frac{z}{1 - z_2}(1 - z_2 - z)Q^2\right]}\\
\frac{\dd \sigma_{12}^{L}}{\dd^2 \bq\, \dd y_2} =& 
\frac{-2e^2g^2Q^2N_c^2 z_2 (1 - z_2)^2}{(2\pi)^5} 
\int \dd^6 \bx\big[S_{2 2^\p}-S_{12}-S_{1 2^\p}+1\big] K_0(|\bx_{1 2^\p}|Q_2) \, 
e^{i\bq\cdot(\bx_2^\p-\bx_2)}
\nonumber\\
&\int_0^{z_2} \frac{\dd z}{z} \left[z_2^2 + (z_2-z)^2\right]
\int \dtwo{\bk_2} \int \dtwo{\bk_1}  \frac{e^{i\bk_1\cdot(\bx_1-\bx_2)}}
{\big[\bk_1^2+Q_2^2\big]\left[ \left(\bk_2-\frac{z}{z_2}\bk_1\right)^2
+ \frac{z (z_2 - z)}{(1 - z_2) z_2^2}\bk_1^2 
+\frac{z}{z_2}(z_2 - z)Q^2\right]}\\
\frac{\dd \sigma_{13(1)}^{L}}{\dd^2 \bq\, \dd y_2} =&  
\frac{2 e^2g^2 Q^2 N_c^2 z_2^2(1-z_2)}{(2\pi)^6}\int_{0}^{z_2} \dd z (1-z_2+z)(z_2-z) 
\int \dd^8\bx [S_{12}S_{1^\p 2^\p} - S_{12} - S_{1^\p 2^\p} +1]e^{i\bq\cdot\bx_{2^\p 2}}  \nonumber \\
&K_0\left(|\bx_{12}|Q\sqrt{(1-z_2+z)(z_2-z)}\right)K_0(|\bx_{1^\p 2^\p}|Q_2)\int \dtwo{\bk}e^{i\bk\cdot\bx_{21}} \nonumber \\
&\int\dtwo{\bp}\Bigg[ \frac{\frac{z_2(z_2-z)}{z}}{(z_2\bk-z\bq)^2} +z_2 z\frac{  
\left(\bk-\bq+\frac{(z_2-z)}{1-z_2}\bp\right)\cdot\left(\bk+\bp-\frac{(1-z_2+z)}{z_2}\bq\right)}
{(1-z_2+z)(z_2\bk-(1-z_2)\bq)^2)\left[\frac{(z_2\bk-(1-z_2)\bq)^2}{z_2 (z_2-z)}-\frac{((1-z_2)\bk-z\bp)^2}
{(1-z_2)(1-z_2+z)}\right]} \Bigg]  e^{i\bp\cdot\bx_{1^\p 1}}\\
\frac{\dd \sigma_{13(2)}^{L}}{\dd^2 \bq\, \dd y_2} =&  \frac{2 e^2g^2 Q^2 N_c^2 z_2^2(1-z_2)}{(2\pi)^6}\int_{0}^{1-z_2} \dd z (1-z_2-z)(z_2+z) \int \dd^8\bx [S_{12}S_{1^\p 2^\p} - S_{12} - S_{1^\p 2^\p} +1] e^{i\bq\cdot\bx_{2^\p 2}}  \nonumber \\
&K_0\left(|\bx_{12}|Q_1\right)K_0(|\bx_{1^\p 2^\p}|Q_2)\int \dtwo{\bk}e^{i\bk\cdot\bx_{12}}\nonumber \\
&\int\dtwo{\bp}\Bigg[\frac{ \frac{(1-z_2)(1-z_2-z)}{z}}{((1-z_2)\bk-z\bp)^2}+(1-z_2) z\frac{\left(\bk-\bp+\frac{(1-z_2-z)}{z_2}\bq\right)\cdot\left(\bk+\bq-\frac{(z_2+z)}{1-z_2}\bp\right)}{(z_2+z)((1-z_2)\bk-z\bp)^2\left[\frac{((1-z_2)\bk-z\bp)^2}{(1-z_2)(1-z_2-z)}-\frac{(z_2\bk-z\bq)^2}{z_2 (z_2+z)}\right]}\Bigg]e^{i\bp\cdot\bx_{1^\p 1}}.
\end{align}

\begin{align}
\frac{\dd \sigma_{14(1)}^{ L}}{\dd^2 \bq\, \dd y_2 } =& 
\frac{e^2 g^2 Q^2 N_c^2 z_2^2 (1 - z_2)}{(2\pi)^5} 
\int \dd^6 \bx [ S_{2 2^\p} - S_{12} - S_{1 2^\p} + 1] K_0(|\bx_{1 2^\p}|Q_2) 
e^{i\bq\cdot(\bx_2^\p - \bx_2)} 
\nonumber \\
& \int_0^{1 - z_2} \frac{\dd z}{z} \dtwo{\bk_1}\dtwo{\bk_2} e^{i\bk_2\cdot\bx_{12}} 
\Bigg[ \frac{z_2 (1 - z_2) +(1 - z_2 - z)(z_2 + z)}
{\Big[\bk_2^2+Q_2^2\Big] 
\left[\left(\bk_1-\frac{1 - z_2 - z}{1 - z_2}\bk_2\right)^2
+\frac{z(1 - z_2 - z)}{z_2 (1 - z_2)^2} \bk_2^2 + \frac{z}{1 - z_2}(1 - z_2 - z)Q^2\right]} \nonumber \\
& + \frac{\frac{(1 - z_2 -z)( z_2  + z)}{z_2(1-z_2)} 
((z_2+z)(1-z_2-z)+z_2(1-z_2)))}
{\Big[\bk_1^2 + (1 - z_2 -z)(z_2  + z)Q^2\Big] 
\left[\left(\bk_1-\frac{1 - z_2 -z}{1 - z_2 }\bk_2\right)^2 
+\frac{z(1 - z_2 -z)}{z_2  (1 - z_2)^2} \bk_2^2 + \frac{z}{1 - z_2 }(1 - z_2 -z)Q^2\right]} \nonumber\\
&- \frac{\frac{z(1 - z_2 -z)}{1 - z_2 } \left[z_2  (1 - z_2 ) + (1 - z_2 -z)(z_2  + z)\right]Q^2}
{\Big[\bk_1^2 + (1 - z_2 -z)(z_2  + z)Q^2\Big]
\Big[\bk_2^2 + Q_2^2\Big] 
\left[\left(\bk_1-\frac{1 - z_2 -z}{1 - z_2 }\bk_2\right)^2
+\frac{z(1 - z_2 -z)}{z_2 (1 - z_2)^2} \bk_2^2 + \frac{z}{1 - z_2 }(1 - z_2 -z)Q^2\right]} \nonumber \\
&- \frac{z \left[(1 - z_2) (1 - z_2 -z) + z_2  (z_2  + z)\right]}
{\Big[\bk_1^2 + (1 - z_2 -z)(z_2 + z)Q^2\Big]
\Big[\bk_2^2+Q_2^2\Big]}  \Bigg] \\
\frac{\dd \sigma_{14(2)}^{ L}}{\dd^2 \bq\, \dd y_2 } =& 
\frac{e^2 g^2 Q^2 N_c^2 z_2^2 (1 - z_2)}{(2\pi)^5} 
\int \dd^6 \bx [ S_{2 2^\p} - S_{12} - S_{1 2^\p } + 1] K_0(|\bx_{1 2^\p }|Q_2) 
e^{i\bq\cdot(\bx_2^\p - \bx_2)} 
\nonumber \\
& \int_0^{z_2} \frac{\dd z}{z} \dtwo{\bk_1}\dtwo{\bk_2} e^{i\bk_2\cdot\bx_{12}} 
\Bigg[ \frac{z_2 (1 - z_2) +(1 - z_2 + z)(z_2 - z)}
{\Big[\bk_2^2+Q_2^2\Big] 
\left[\left(\bk_1-\frac{z_2 - z}{z_2}\bk_2\right)^2
+\frac{z(z_2 - z)}{(1 - z_2) z_2^2} \bk_2^2 
+ \frac{z}{z_2}(z_2-z)Q^2\right]} \nonumber \\
& + \frac{\frac{(1 - z_2 + z)(z_2 - z)}{z_2 (1 - z_2)} 
\left[z_2 (1 - z_2) + (1 - z_2 + z)(z_2 - z)\right]}
{\Big[\bk_1^2 + (1 - z_2 + z)(z_2 - z)Q^2\Big] 
\left[\left(\bk_1-\frac{z_2 - z}{z_2}\bk_2\right)^2
+\frac{z(z_2 - z)}{(1 - z_2) z_2^2} \bk_2^2 
+ \frac{z}{z_2}(z_2 - z)Q^2\right]} \nonumber\\
&- \frac{\frac{z (z_2 - z)}{z_2} \left[z_2 (1 - z_2) + (1 - z_2 + z)(z_2 - z)\right]Q^2}
{\Big[\bk_1^2 + (1 - z_2 + z)(z_2 - z)Q^2\Big]
\Big[\bk_2^2 + Q_2^2\Big] 
\left[\left(\bk_1-\frac{z_2 - z}{z_2}\bk_2\right)^2
+\frac{z(z_2 - z)}{(1 - z_2) z_2^2} \bk_2^2 
+ \frac{z}{z_2}(z_2 - z)Q^2\right]} \nonumber \\
&- \frac{z \left[z_2 (z_2 - z) + (1 - z_2) (1 - z_2 + z)\right]}
{\Big[\bk_1^2 + (1 - z_2 + z)(z_2 - z)Q^2\Big]
\Big[\bk_2^2+Q_2^2\Big]}  \Bigg]\label{142}
\end{align}
where we have 
 $\bx_1^\p \equiv \bx_1 + \frac{z}{1 - z_2}(\bx_3 - \bx_1)$ in $\sigma_{1\times 2}$. Here we have also defined 

\begin{align}
\Delta^{(3)}_{ij} = \frac{\bx_{3i}\cdot\bx_{3j}}{\bx_{3i}^2 \bx_{3j}^2}.
\end{align}

\noindent We also note that since we have integrated over $z_1$ the definition of $Q_2$ remains the same but $Q_1$ (which is still used in some expressions) has now changed.

\begin{align}
Q_2 = Q\sqrt{z_2(1-z_2)}, \,\,\,\,\,\,\,\, Q_1 = Q\sqrt{(1-z_2-z)(z_2+z)}.
\end{align}

\noindent We have also used a shorthand notation for the coordinate dependence in some of the Bessel functions.

\begin{align}
X &= \sqrt{(1-z_2-z) z_2 \bx_{12}^2 + (1-z_2-z) z \bx_{13}^2 + z_2 z \bx_{23}^2},\nonumber \\
X_6 &= \sqrt{(1-z_2)(z_2-z)\bx_{12}^2 + (1-z_2) z\,\bx_{13}^2 + z(z_2-z)\bx_{23}^2}.
\end{align}

\noindent $X^\prime$ is the same as $X$ but with primed coordinates (except for $\bx_3$ which is never primed). $X^\prime_1$ is the same as $X^\prime$ but $\bx_1^\p$ has become unprimed.

\begin{figure}[H]
\centering
\includegraphics[width=160mm]{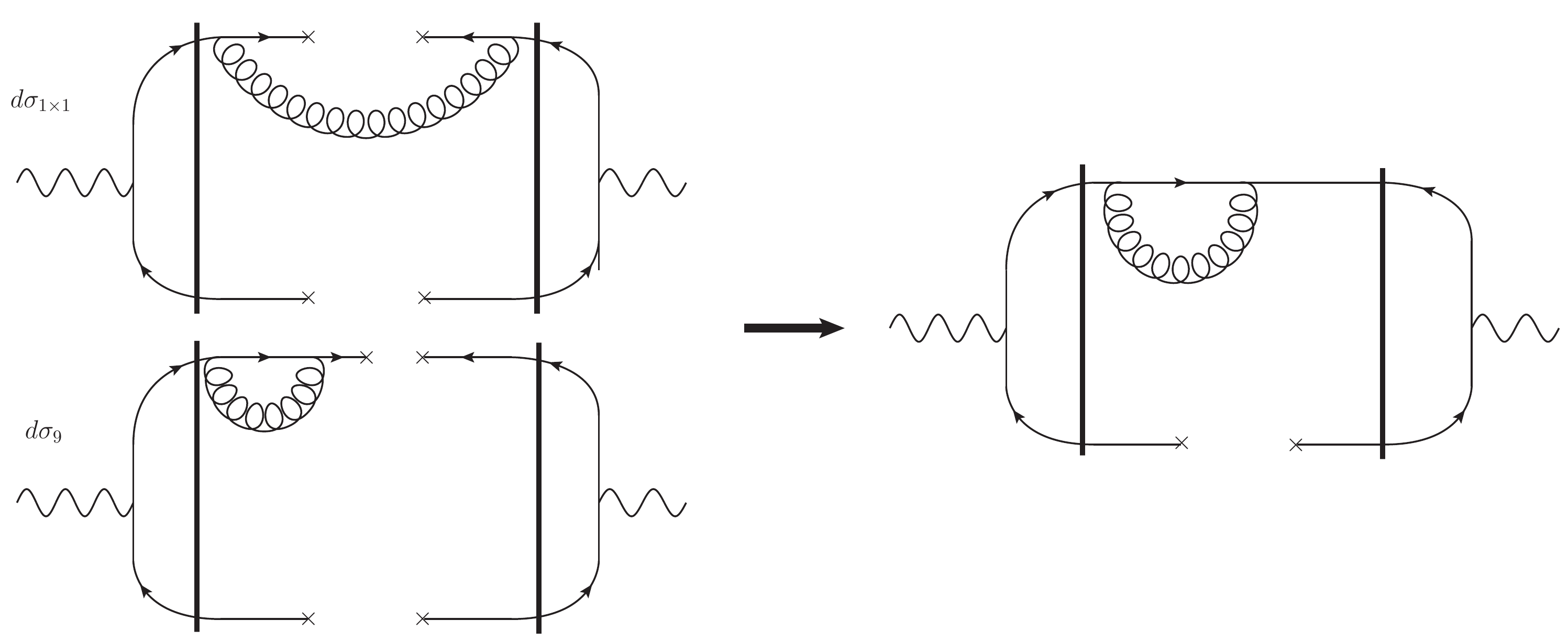}
\caption{When one integrates out the final state quark, the real correction $\sigma_{1\times 1}$ and the virtual correction $\sigma_9$ become the same diagram. Here the `x' at the end of a solid line indicates produced quarks and antiquarks. }\label{fig:cutdiag1}
\end{figure}

\begin{figure}[H]
\centering
\includegraphics[width=70mm]{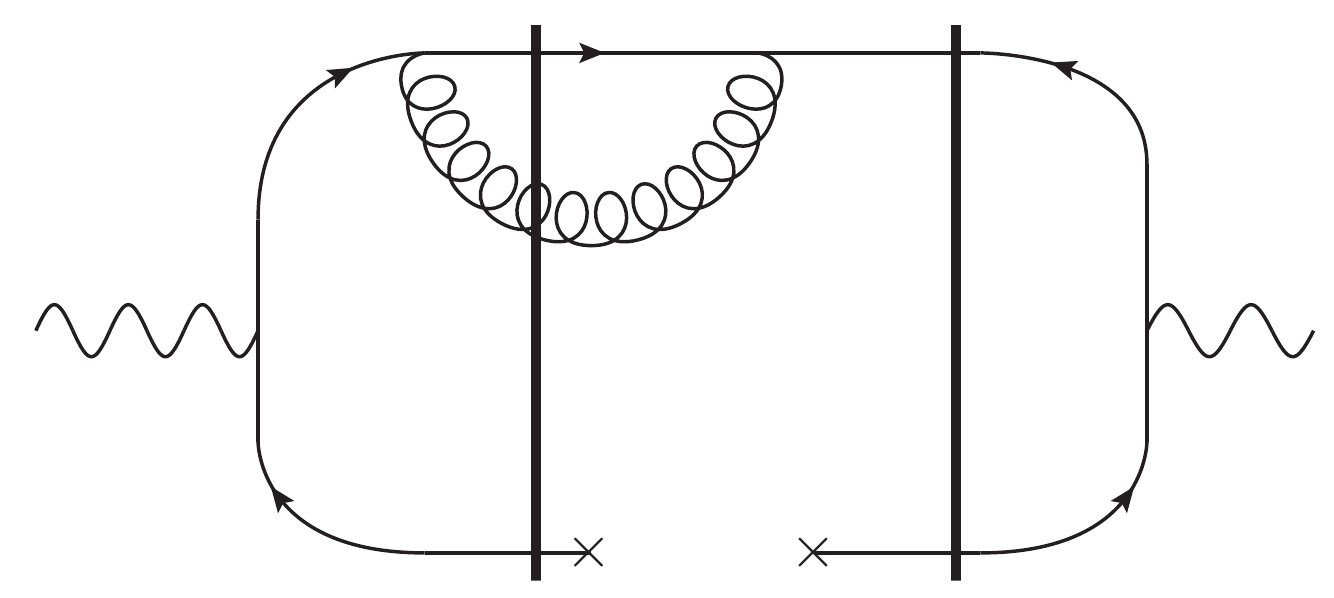} \includegraphics[width=70mm]{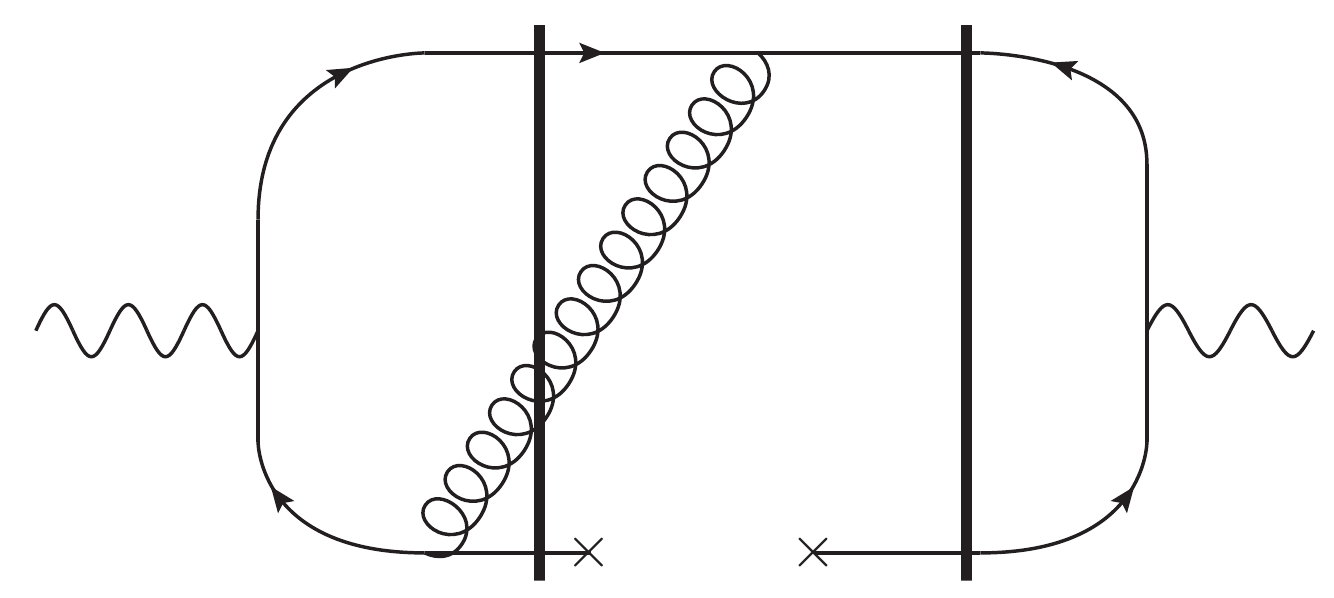}
\caption{Here the left diagram is $\sigma_{1\times3}$ and $\sigma_5$, and the right diagram is $\sigma_{1\times4}$ and $\sigma_7$. The `x' indicates the produced antiquark. }\label{fig:cutdiags}
\end{figure}

 These expressions provide the formal results for the one-loop corrections to single inclusive 
 hadron production. Looking at the results (Eq. \ref{2x2} - \ref{142}) one can see that some corrections appear to be missing. In particular, we have not written $\sigma_{1\times 1}, \sigma_{1\times 3},\sigma_{1\times 4}, \sigma_5, \sigma_7,$ and $\sigma_9$. This is because $\sigma_{1\times 1}$ exactly cancels $\sigma_9$ ($\sigma_9$ gets an extra factor of $2$ due to it being a cross term). Similarly, $\sigma_{1\times 3}$ cancels $\sigma_5$, and also $\sigma_{1\times 4}$ cancels $\sigma_7$. We include the expressions for these in appendix \ref{appcancel} for completeness. To understand this cancellation, one can draw these corrections in cut diagram notation. In Fig. \ref{fig:cutdiag1} we show that $\sigma_{1\times 1}$ and $\sigma_9$ become diagrammatically identical when one integrates the final state quark. Similarly, one finds that $\sigma_{1\times 3}$ becomes the same diagram as $\sigma_5$, and $\sigma_{1\times 4}$ becomes the same as $\sigma_7$ (see Fig. \ref{fig:cutdiags}).

\noindent Therefore since the expressions corresponding to these diagrams differ mathematically only by a sign, they all cancel each other completely. 

\begin{align}
&\dd\sigma_{1\times 1}+2\dd\sigma_9 = 0, \nonumber \\
&\dd\sigma_{1\times 3}+\dd\sigma_5 = 0, \nonumber \\
&\dd\sigma_{1\times 4} +\dd\sigma_7 = 0. \label{fullcancel}
\end{align}

\noindent Note that the factor of 2 on $\dd\sigma_9$ comes due to the fact that it's a cross term and therefore gets double counted relative to $\dd\sigma_{1\times 1}$. So it is justified to ignore these corrections (see appendix \ref{appcancel} for the expressions). In the remaining terms, one finds divergences which must either be canceled or absorbed into the renormalization of physical parameters. This is the topic of the next section.
 
 \section{Divergences}
 
 As in the case of dihadron production in \cite{Bergabo:2022tcu} there are 
 four categories of divergences;
 UV, soft, rapidity and collinear. The cancellation of UV and soft divergences 
 proceeds as follows. UV divergences appear as $\bx_3 \to \bx_2$ in $\sigma_6$, as $\bx_3 \to \bx_1$ in $\sigma_{3\times 3}$, as $\bk\to\infty$ in $\sigma_{10}$, as $\bk_2 \to \infty$ in $\sigma_{11}$ and $\sigma_{12}$, and as $\bk_1 \to \infty$ in $\sigma_{14(1)}$ and $\sigma_{14(2)}$. We find that these UV divergences all cancel according to Eq. \ref{UV}. Note that all other UV divergent terms were already canceled in the first two lines of Eq. \ref{fullcancel}. 
 
\begin{align}
&\left[\dd \sigma_{6}+\dd \sigma_{12} \right]_{\text{UV}} = 0, \nonumber \\
&\left[\dd \sigma_{3\times 3}+\dd \sigma_{11}+\dd\sigma_{10}+\dd\sigma_{14(1)}+\dd\sigma_{14(2)}\right]_{\text{UV}}  = 0. \label{UV}
\end{align}

Soft divergences occur when $\bk$ and $z$ both go to zero (in coordinate space this becomes $\bx_3 \to \infty$ and $z \to 0$). The cancellation of soft divergences in all remaining terms proceeds identically to that in dihadron production~\cite{Bergabo:2022tcu} with 
\begin{align}
&\left[\dd \sigma_{2\times 2} + 2\,\dd \sigma_{10}\right]_{\text{soft}} = 0, \nonumber \\
&\left[\dd \sigma_{1\times 2} + \dd\sigma_{13(1)}+\dd\sigma_{13(2)} \right]_{\text{soft}}= 0, \nonumber \\
&\left[\dd \sigma_{3\times 3} + \dd \sigma_{4\times 4} + 2\,\dd \sigma_{3\times 4}\right]_{\text{soft}} = 0,\nonumber\\
&\left[\dd \sigma_{2\times 3} + \dd \sigma_{2\times 4}\right]_{\text{soft}} = 0,\nonumber\\
&\left[\dd \sigma_{6}+\dd\sigma_{8}\right]_{\text{soft}} = 0, \nonumber \\
&\left[\dd \sigma_{11} + \dd \sigma_{14(1)}\right]_{\text{soft}} = 0,  \nonumber \\
&\left[\dd \sigma_{12} + \dd \sigma_{14(2)}\right]_{\text{soft}} = 0 .
\end{align}

Rapidity divergences appear when $z \rightarrow 0$ at finite (non-zero) transverse momentum. 
To isolate those it is customary to introduce a rapidity factorization scale $z_f$
and write the $z$ integral as     
\begin{align}
\int_0^1 \frac{\dd z}{z} f (z) =  \left\{\int_0^{z_f} \frac{\dd z}{z}  + 
\int_{z_f}^1 \frac{\dd z}{z} \, \right\} \, f (z).
\end{align}
so that the rapidity divergences will come from the first integral while the cross 
section in the second integral will contain no rapidity divergence. Therefore we first 
focus on the first integration region containing the rapidity divergence. As all soft 
divergences have already been canceled only terms of the form 
$ \frac{\bx_{i j}^2}{\bx_{3 i}^2\bx_{3 j}^2}$ with $i \neq j$ remain (see \cite{Bergabo:2022tcu} 
for the explicit expressions). These can be added to give

\begin{align}
\frac{\dd \sigma^{L}}{\dd^2 \bq \, \dd y_2} =& 
\frac{4 e^2 Q^2 N_c z_2^3 (1-z_2)^2 }{(2\pi)^7} \, 
\int \dd^6 \bx K_0(|\bb{x}_{12}|Q_1)K_0(|\bb{x}_{1 2^\prime}|Q_1) \nonumber \\
&
\bigg\{
\frac{N_c \, \alpha_s}{2 \pi^2} \, 
 \int_0^{z_f}\frac{\dd z}{z} \int \dd^{2} \bx_3 
 \bigg[
 \frac{\bx_{2 2^\p}^2}{\bx_{3 2}^2\bx_{3 2^\p}^2}\left(S_{2 3} S_{2^\p 3} - S_{2 2^\p}\right)
 - \frac{\bx_{1 2}^2}{\bx_{3 1}^2\bx_{3 2}^2} \left(S_{1 3} S_{3 2} - S_{1 2}\right) 
 - \frac{\bx_{1 2^\p}^2}{\bx_{3 1}^2\bx_{3 2^\p}^2} \left(S_{1 3} S_{2^\p 3} - S_{1 2^\p}\right)
 \bigg]
 \bigg\}.\\
\end{align}
Comparing this to the LO result in Eq. \ref{LOdsig}
it is clear that the terms inside the curly bracket
correspond to the BK/JIMWLK evolution~
\cite{Balitsky:1995ub,Kovchegov:1999yj,Jalilian-Marian:1997qno,Jalilian-Marian:1997jhx,Jalilian-Marian:1997ubg} 
of the dipoles that appear in the LO cross section.
 As the contribution of the second term in the 
$z$ integral contains no rapidity divergence this shows that all rapidity divergences can
be absorbed into BK/JIMWLK evolution of the LO cross section. The contribution of the 
$\int_{z_f}^1\dd z$ region is now free of divergences (after absorbing the collinear divergences 
into scale dependent fragmentation functions done in the following pages) and constitute the NLO
correction to the LO result.

Collinear divergences are identical to the case of dihadrons and were treated in full detail 
in~\cite{Bergabo:2022tcu}, therefore here we will skip some details. When we 
integrate out quarks there remains collinear divergences involving the antiquark when 
its transverse momentum $\bq$ becomes parallel to the loop transverse momentum 
$\bk$ ($\theta \rightarrow 0$ with $\cos\theta \equiv \frac{\bq \cdot\bk}{|\bq||\bk|}$ at finite transverse momenta). 
One can write the hadronic cross section in terms of the partonic cross section as follows,

\begin{align}
\frac{\dd \sigma^{\gamma^* A\to h X}}{\dd^2 \bq_h \, \dd y_2} = 2\int_0^1 \frac{\dd z_h}{z_h^2} \frac{\dd \sigma^{\gamma^* A \to \barq X}}{\dd^2 \bq \,\dd y_2} D_{h/\barq}(z_h).\label{hadronic}
\end{align}

\noindent Here we have included only the case where the quark is integrated out and the antiquark with momentum $q$ fragments into a hadron $h$ with with momentum $z_h q$. As mentioned earlier, the opposite case (antiquark integrated out and the quark fragments into a hadron) is mathematically the same after a relabeling of some variables. Therefore we account for the opposite case with the overall factor of 2 above. The renormalized fragmentation function $D_{h/\barq}(z_h, \mu^2) $ can be written in terms of the \emph{bare} fragmentation function $D^0_{h/\barq}$ and higher order corrections.

\begin{align}
D_{h/\barq}(z_h, \mu^2) = D^{0}_{h/\barq}(z_h) + \mathcal{O}(\alpha_s)
\end{align}

\noindent The $\mathcal{O}(\alpha_s)$ corrections are expected to come from $\sigma_{2\times 2}$ (Eq. \ref{2x2}) and $\sigma_{10}$ (Eq. \ref{10}). Using Eq. \ref{hadronic} with the leading order partonic cross section (longitudinal part of Eq. \ref{LOdsig}) and the two relevant corrections (in momentum space), we can write 

\begin{align}
\frac{\dd \sigma^{\gamma^* A \to h X}_{\text{LO}+2\times2 +10}}{\dd^2 \bq_h\, \dd y_2} =& \int_0^1 \frac{\dd z_h}{z_h^2} D^0_{h/\barq}(z_h) \frac{2e^2 Q^2 N_c}{(2\pi)^5} \int \dd^6 \bx[S_{22^\p}-S_{12}-S_{12^\p}+1] e^{i\bq\cdot\bx_{2^\p 2}} K_0(|\bx_{12}|Q_2)K_0(|\bx_{12^\p}|Q_2) \nonumber \\
&\Bigg[ 4z_2^3(1-z_2)^2 +\frac{2g^2 N_c}{(2\pi)} \int\frac{\dd z}{z} \frac{(1-z_2-z)^2(z_2+z)^2[z_2^2+(z_2+z)^2]}{z_2} \int\dtwo{\bk} \frac{e^{i\bk\cdot\bx_{2^\p 2}}}{\left(\bk-\frac{z}{z_2}\bq\right)^2} \nonumber \\
&-\frac{2g^2N_c}{(2\pi)}\int_0^{z_2}\frac{\dd z}{z} z_2(1-z_2)^2[z_2^2+(z_2-z)^2] \int\dtwo{\bk}\frac{1}{\left(\bk-\frac{z}{z_2}\bq\right)^2}\Bigg].
\end{align}

\noindent Here the first term in the square brackets is the leading order contribution, the second term is the real correction $\sigma_{2\times 2}$, and the last term is the virtual correction $\sigma_{10}$ (doubled here since it's a cross term). Next we follow identical steps to what was done for dihadrons and rewrite this expression using $g^2 = 4\pi \alpha_s$ and relax the large-$N_c$ approximation inside the square brackets taking $N_c \to 2C_F$. We also define a new variable $\xi$ for the $z$ integration in both the real and virtual corrections. For the real correction $\xi = z_2/(z_2+z)$ and for the virtual correction $\xi = (z_2-z)/z_2$. 

\begin{align}
\frac{\dd \sigma^{\gamma^* A \to h X}_{\text{LO}+2\times2 +10}}{\dd^2 \bq_h\, \dd y_2} =& \int_0^1 \frac{\dd z_h}{z_h^2} D^0_{h/\barq}(z_h) \frac{8e^2 Q^2 N_cz_2^3}{(2\pi)^5} \int \dd^6 \bx[S_{22^\p}-S_{12}-S_{12^\p}+1] e^{i\bq\cdot\bx_{2^\p 2}} K_0(|\bx_{12}|Q_2)K_0(|\bx_{12^\p}|Q_2) \nonumber \\
&\Bigg[ (1-z_2)^2 +2 \alpha_s C_F \int\frac{\dd \xi}{\xi^5}\frac{(1-z_2/\xi)^2 (1+\xi^2)}{(1-\xi)}\int\dtwo{\bk} \frac{e^{i\bk\cdot\bx_{2^\p 2}}}{\left(\bk-\frac{(1-\xi)}{\xi}\bq\right)^2} \nonumber \\
&-2(1-z_2)^2 \alpha_s C_F\int_0^{1}\dd \xi\frac{(1+\xi^2)}{(1-\xi)} \int\dtwo{\bk}\frac{1}{\left(\bk-(1-\xi)\bq\right)^2}\Bigg].
\end{align}

\noindent Now, we want to regulate the divergences in the two integrals over $\bk$. To do so we follow \cite{Bergabo:2022tcu} and write 

\begin{align}
& \int \dtwo{\bk} \frac{e^{i\bk\cdot(\bx_2^\p-\bx_2)}}{\left(\bk-\frac{(1-\xi)}{\xi}\bq\right)^2} \to  \frac{ e^{i\frac{(1-\xi)}{\xi}\bq\cdot(\bx_2^\p - \bx_2)}}{2\pi} \left[ \frac{1}{\epsilon} -\log\left(\pi e^{\gamma_E}\mu |\bx_2^\p-\bx_2|\right)\right] + \mathcal{O}(\epsilon),\,\,\,\,\, \epsilon = d-2 > 0. \label{realdr}
\end{align}

\begin{align}
\left[\int \dtwo{\bk}\frac{1}{\bk^2}\right]_{IR} = \frac{1}{2\pi} \left[ -\frac{1}{\epsilon_{IR}} -\log\left(e^{\gamma_E} \pi \mu |\bx_2^\p - \bx_2|\right)\right] + \mathcal{O}(\epsilon). \label{eir}
\end{align}

\begin{align}
\left[\int \dtwo{\bk}\frac{1}{\bk^2}\right]_{UV} = \frac{1}{2\pi} \left[ \frac{1}{\epsilon_{UV}}  + \log\left(e^{\gamma_E} \pi \mu |\bx_2^\p - \bx_2|\right)\right] + \mathcal{O}(\epsilon).\label{euv}
\end{align}

\noindent Here Eq. \ref{realdr} is used for the real integral. We also make the approximation $\xi \approx 1$ inside the exponential on the right side to ignore it, this we motivate by noting that there is a $1-\xi$ in the denominator so the integral is dominated by the region $\xi \approx 1$. Eq. \ref{eir} and \ref{euv} are used for the virtual correction, after a shift on the $\bk$ integral. This shift makes the collinear divergence look infrared, but this is distinct from the soft divergences discussed earlier and is in fact the collinear divergence. The UV divergence is canceled against other virtual corrections (Eq. \ref{UV}). Therefore what remains here is the finite part of Eq.\ref{euv} which remains as part of the finite NLO corrections, and Eq. \ref{eir} which can now be added to the real correction by setting $\epsilon_{IR} = -\epsilon$. 

\begin{align}
&\frac{\dd \sigma^{\gamma^* A \to h X}_{\text{LO}+2\times2 +10}}{\dd^2 \bq_h\, \dd y_2} = \int_0^1 \frac{\dd z_h}{z_h^2} D^0_{h/\barq}(z_h) \frac{8e^2 Q^2 N_c z_2^3}{(2\pi)^5} \int \dd^6 \bx[S_{22^\p}-S_{12}-S_{12^\p}+1] e^{i\bq\cdot\bx_{2^\p 2}} K_0(|\bx_{12}|Q_2)K_0(|\bx_{12^\p}|Q_2) \nonumber \\
&\Bigg[(1-z_2)^2 +\left\{\frac{ \alpha_s C_F}{\pi} \int\frac{\dd \xi}{\xi^5}\frac{ (1-z_2/\xi)^2(1+\xi^2)}{(1-\xi)}-\frac{ (1-z_2)^2\alpha_s C_F}{\pi}\int_0^{1}\dd \xi\frac{(1+\xi^2)}{(1-\xi)}\right\} \left[ \frac{1}{\epsilon} -\log\left(\pi e^{\gamma_E}\mu |\bx_2^\p-\bx_2|\right)\right] \Bigg]
\end{align}

\noindent Next, let's note that $z_2$ is no longer an independent variable, it can be written in terms of $z_h$ using 

\begin{align}
 z_2 = \frac{q_h^+}{z_h l^+}.
\end{align}

\noindent Writing all the $z_2$'s this way, and rearranging the result we have

\begin{align}
&\frac{\dd \sigma^{\gamma^* A \to h X}_{\text{LO}+2\times2 +10}}{\dd^2 \bq_h\, \dd y_2} = \int_0^1\dd z_h  \frac{8e^2 Q^2 N_c (q_h^+)^3}{(2\pi)^5(l^+)^3} \int \dd^6 \bx[S_{22^\p}-S_{12}-S_{12^\p}+1] e^{i\bq\cdot\bx_{2^\p 2}} K_0(|\bx_{12}|Q_2)K_0(|\bx_{12^\p}|Q_2) \nonumber \\
&\Bigg[\frac{\left(1-\frac{q_h^+}{z_h l^+}\right)^2}{z_h^5}D^0_{h/\barq}(z_h) +\left\{\frac{ \alpha_s C_F}{\pi} \int\frac{\dd \xi}{\xi^5}\frac{ (1-\frac{q_h^+}{z_h l^+ \xi})^2(1+\xi^2)}{(1-\xi)}\frac{D^0_{h/\barq}(z_h)}{z_h^5}-\frac{ (1-\frac{q_h^+}{z_h l^+})^2\alpha_s C_F}{\pi}\int_0^{1}\dd \xi\frac{(1+\xi^2)}{(1-\xi)}\frac{D^0_{h/\barq}(z_h)}{z_h^5}\right\}\nonumber \\
&\times \left[ \frac{1}{\epsilon} -\log\left(\pi e^{\gamma_E}\mu |\bx_2^\p-\bx_2|\right)\right] \Bigg].
\end{align}

\noindent Now that all the $z_h$ dependence is explicit, let's perform a substitution on the $z_h$ integral only in the real correction term (first term in the curly brackets). We'll define $z_h^\p = \xi z_h$, and once the substitution is complete we'll remove the prime. This yields

\begin{align}
&\frac{\dd \sigma^{\gamma^* A \to h X}_{\text{LO}+2\times2 +10}}{\dd^2 \bq_h\, \dd y_2} = \int_0^1\frac{\dd z_h}{z_h^2}  \frac{8e^2 Q^2 N_c z_2^3(1-z_2)^2}{(2\pi)^5} \int \dd^6 \bx[S_{22^\p}-S_{12}-S_{12^\p}+1] e^{i\bq\cdot\bx_{2^\p 2}} K_0(|\bx_{12}|Q_2)K_0(|\bx_{12^\p}|Q_2) \nonumber \\
&\Bigg[D^0_{h/\barq}(z_h) +\left\{\frac{ \alpha_s C_F}{\pi} \int_{z_h}^1\frac{\dd \xi}{\xi}\frac{(1+\xi^2)}{(1-\xi)}D^0_{h/\barq}(z_h/\xi)-\frac{\alpha_s C_F}{\pi}\int_0^{1}\dd \xi\frac{(1+\xi^2)}{(1-\xi)}D^0_{h/\barq}(z_h)\right\} \left[ \frac{1}{\epsilon} -\log\left(\pi e^{\gamma_E}\mu |\bx_2^\p-\bx_2|\right)\right] \Bigg].
\end{align}

\noindent Here we have written things back in terms of $z_2$ for aesthetics. Finally, the real and virtual corrections can be combined using the antiquark-antiquark splitting function $\mathcal{P}_{\barq \barq}$ defined as

\begin{align}
\mathcal{P}_{\barq \barq}(\xi) = C_F\left[\frac{ (1+\xi^2)}{(1-\xi)_+}+\frac32 \delta(1-\xi)\right]
\end{align}

\begin{align}
&\frac{\dd \sigma^{\gamma^* A \to h X}_{\text{LO}+2\times2 +10}}{\dd^2 \bq_h\, \dd y_2} = \int_0^1\frac{\dd z_h}{z_h^2}  \frac{8e^2 Q^2 N_c z_2^3(1-z_2)^2}{(2\pi)^5} \int \dd^6 \bx[S_{22^\p}-S_{12}-S_{12^\p}+1] e^{i\bq\cdot\bx_{2^\p 2}} K_0(|\bx_{12}|Q_2)K_0(|\bx_{12^\p}|Q_2) \nonumber \\
&\int_{z_h}^1 \frac{\dd \xi}{\xi} D^0_{h/\barq}(z_h/\xi)\Bigg[\delta(1-\xi) +\frac{ \alpha_s}{\pi}P_{\barq \barq}(\xi) \left[ \frac{1}{\epsilon} -\log\left(\pi e^{\gamma_E}\mu |\bx_2^\p-\bx_2|\right)\right] \Bigg].
\end{align}

\noindent For more details on these calculations, see \cite{Bergabo:2022tcu}. So, we are now able to define the DGLAP evolved fragmentation function 

\begin{align}
D_{h/\barq}(z_h,\mu^2) = \int_{z_h}^1 \frac{\dd \xi}{\xi} D^0_{h/\barq}(z_h/\xi)\Bigg[\delta(1-\xi) +\frac{ \alpha_s}{\pi}P_{\barq \barq}(\xi) \left[ \frac{1}{\epsilon} -\log\left(\pi e^{\gamma_E}\mu |\bx_2^\p-\bx_2|\right)\right] \Bigg],
\end{align}

\noindent in terms of which our result becomes

\begin{align}
&\frac{\dd \sigma^{\gamma^* A \to h X}_{\text{LO}+2\times2 +10}}{\dd^2 \bq_h\, \dd y_2} = \int_0^1\frac{\dd z_h}{z_h^2}  \frac{8e^2 Q^2 N_c z_2^3(1-z_2)^2}{(2\pi)^5} \int \dd^6 \bx[S_{22^\p}-S_{12}-S_{12^\p}+1] e^{i\bq\cdot\bx_{2^\p 2}} K_0(|\bx_{12}|Q_2)K_0(|\bx_{12^\p}|Q_2) D_{h/\barq}(z_h).
\end{align}

Thus we have shown that all divergences appearing in our next-to-leading order results are either canceled or absorbed into evolution of dipoles and fragmentation functions. The full result for the single inclusive cross section at next-to-leading order can be written schematically as

\begin{align}
\dd \sigma^{\gamma^* A \to h X} = \dd \sigma_{LO}\otimes \text{JIMWLK} + \dd \sigma_{LO} \otimes D_{h/\barq}(z_h,\mu^2) + \dd \sigma_{NLO}^{\text{finite}}.
\end{align}

\noindent Here we imply the presence of a bare fragmentation function $D^0_{h/\barq}(z_h)$ in the first and last terms. 

In summary we have derived the next to leading order corrections to single inclusive hadron production in the forward rapidity region in DIS at small $x$. We have shown that all divergences either cancel among various terms or can be absorbed into BK/JIMWLK evolution of dipoles and DGLAP evolution of the parton-hadron fragmentation function.



\paragraph{Acknowledgement:}We gratefully acknowledge support from the DOE Office of Nuclear Physics through Grant No. DE-SC0002307 and by PSC-CUNY through grant No. 63158-0051. We would like to thank Y. Kovchegov, C. Marquet, B. Xiao and F. Yuan for helpful discussions.

\begin{appendices}
\section{Cancellations}\label{appcancel}

Here we list the corrections that all cancel each other exactly according to Eq. \ref{fullcancel} and can therefore be neglected from the results.

\begin{align}
\frac{\dd \sigma^{L}_{1\times 1}}{\dd^2 \bq\, \dd y_2} = &\frac{2 e^2 g^2  Q^2 N_c^2 
z_2^3 }{(2\pi)^{8}}
\int_0^{1-z_2} \frac{\dd z}{z}\left[ (1 - z_2 - z)^2 + (1 - z_2)^2\right]  \int \dd^{8} \bx \,K_0(|\bx_{1 2}|Q_2)K_0(|\bx_{1 2^\p}|Q_2)
\frac{1}{\bx_{31}^2} \nonumber \\ 
&\left[ S_{2 2^\p} - S_{12} - S_{1 2^\p} + 1\right] 
e^{i\bq\cdot(\bx_2^\prime - \bx_2)}\label{1x1} \\
\nonumber \\
\frac{\dd \sigma^L_{1\times 3}}{\dd^2 \bq\, \dd y_2} =& \frac{-2e^2g^2Q^2 N_c^2 z_2^3}{(2\pi)^{8} } 
\int_0^{1 - z_2} \frac{\dd z}{z} \int \dd^{8} \bx \,
[S_{3 2 2^\p 1^\p} S_{13} - S_{13}S_{23} - S_{1^\p 2^\p} + 1]
\left[(1 - z_2)^2 + (1 - z_2 - z)^2\right] \nonumber \\
&\frac{K_0(Q X)K_0(|\bx_{1^\p 2^\p}|Q_2)}{\bx_{31}^2} 
e^{i\bq\cdot(\bx_2^\p-\bx_2)}
\\
\nonumber\\
\frac{\dd \sigma^{L}_{1\times 4}}{ \dd^2 \bq \, \dd y_2} = & \frac{2e^2g^2Q^2 N_c^2 
z_2^2 }{(2\pi)^{8}} \int_0^{1 - z_2}\frac{\dd z \,(1 - z_2- z)}{z} \int \dd^{8} \bx\,  
[S_{322^\p 1^\p} S_{13} - S_{13}S_{23} - S_{1^\p 2^\p} + 1]
\left[(1 - z_2) z_2 + (1 - z_2 - z)(z_2+z)\right] \nonumber \\
& K_0(QX)  K_0(|\bx_{1^\p2^\p}|Q_2) \Delta^{(3)}_{12} 
e^{i\bq\cdot(\bx_2^\p-\bx_2)} 
\\
\nonumber \\
\frac{\dd \sigma^{L}_5}{\dd^2 \bq \dd y_2} = & \frac{2e^2g^2Q^2 N_c^2 z_2^3}{(2\pi)^{8} } 
\int_0^{1 - z_2} \frac{\dd z}{z} \int \dd^{8} \bx \,
[S_{3 2 2^\p 1^\p} S_{13} - S_{13}S_{23} - S_{1^\p 2^\p} + 1]
\left[(1 - z_2)^2 + (1 - z_2 - z)^2\right] \nonumber \\
&\frac{K_0(Q X)K_0(|\bx_{1^\p 2^\p}|Q_2)}{\bx_{31}^2} 
e^{i\bq\cdot(\bx_2^\p-\bx_2)}
\\
\nonumber \\
\frac{\dd \sigma^{L}_{7}}{ \dd^2 \bq \, \dd y_2} = & \frac{-2e^2g^2Q^2 N_c^2 
z_2^2 }{(2\pi)^{8}} \int_0^{1 - z_2}\frac{\dd z \,(1 - z_2- z)}{z} \int \dd^{8} \bx\,  
[S_{322^\p 1^\p} S_{13} - S_{13}S_{23} - S_{1^\p 2^\p} + 1]
\left[(1 - z_2) z_2 + (1 - z_2 - z)(z_2+z)\right] \nonumber \\
& K_0(QX)  K_0(|\bx_{1^\p2^\p}|Q_2) \Delta^{(3)}_{12} 
e^{i\bq\cdot(\bx_2^\p-\bx_2)} 
\\
\nonumber\\
\frac{\dd \sigma_9^{L}}{\dd^2 \bq \, \dd y_2} = &\frac{-e^2 g^2 Q^2 N_c^2 z_2^3 } 
{(2\pi)^8} \int \dd^8 \bx \big[S_{2 2^\p} - S_{12} - S_{1 2^\p} + 1\big] 
K_0(|\bx_{1 2}|Q_2) K_0(|\bx_{1 2^\p}|Q_2) 
e^{i\bq\cdot(\bx_2^\p-\bx_2)}
\nonumber \\
&\times \int_0^{1 - z_2} \frac{\dd z}{z} [(1 - z_2)^2 + (1 - z_2-z)^2]
\frac{1}{\bx_{31}^2}
\\
\end{align}

\noindent Here  $\bx_1^\p \equiv \bx_1 + \frac{z}{1 - z_2}(\bx_3 - \bx_1)$ in these expressions.
\end{appendices}

\bibliography{mybib}

\begin{thebibliography}{110}
\expandafter\ifx\csname natexlab\endcsname\relax\def\natexlab#1{#1}\fi
\expandafter\ifx\csname bibnamefont\endcsname\relax
  \def\bibnamefont#1{#1}\fi
\expandafter\ifx\csname bibfnamefont\endcsname\relax
  \def\bibfnamefont#1{#1}\fi
\expandafter\ifx\csname citenamefont\endcsname\relax
  \def\citenamefont#1{#1}\fi
\expandafter\ifx\csname url\endcsname\relax
  \def\url#1{\texttt{#1}}\fi
\expandafter\ifx\csname urlprefix\endcsname\relax\def\urlprefix{URL }\fi
\providecommand{\bibinfo}[2]{#2}
\providecommand{\eprint}[2][]{\url{#2}}

\bibitem[{\citenamefont{Gribov et~al.}(1983)\citenamefont{Gribov, Levin, and
  Ryskin}}]{Gribov:1984tu}
\bibinfo{author}{\bibfnamefont{L.~V.} \bibnamefont{Gribov}},
  \bibinfo{author}{\bibfnamefont{E.~M.} \bibnamefont{Levin}}, \bibnamefont{and}
  \bibinfo{author}{\bibfnamefont{M.~G.} \bibnamefont{Ryskin}},
  \bibinfo{journal}{Phys. Rept.} \textbf{\bibinfo{volume}{100}},
  \bibinfo{pages}{1} (\bibinfo{year}{1983}).

\bibitem[{\citenamefont{Mueller and Qiu}(1986)}]{Mueller:1985wy}
\bibinfo{author}{\bibfnamefont{A.~H.} \bibnamefont{Mueller}} \bibnamefont{and}
  \bibinfo{author}{\bibfnamefont{J.-w.} \bibnamefont{Qiu}},
  \bibinfo{journal}{Nucl. Phys. B} \textbf{\bibinfo{volume}{268}},
  \bibinfo{pages}{427} (\bibinfo{year}{1986}).

\bibitem[{\citenamefont{Iancu and Venugopalan}(2003)}]{Iancu:2003xm}
\bibinfo{author}{\bibfnamefont{E.}~\bibnamefont{Iancu}} \bibnamefont{and}
  \bibinfo{author}{\bibfnamefont{R.}~\bibnamefont{Venugopalan}},
  \emph{\bibinfo{title}{{The Color glass condensate and high-energy scattering
  in QCD}}} (\bibinfo{year}{2003}), pp. \bibinfo{pages}{249--3363},
  \eprint{hep-ph/0303204}.

\bibitem[{\citenamefont{Iancu et~al.}(2002)\citenamefont{Iancu, Leonidov, and
  McLerran}}]{Iancu:2002xk}
\bibinfo{author}{\bibfnamefont{E.}~\bibnamefont{Iancu}},
  \bibinfo{author}{\bibfnamefont{A.}~\bibnamefont{Leonidov}}, \bibnamefont{and}
  \bibinfo{author}{\bibfnamefont{L.}~\bibnamefont{McLerran}}, in
  \emph{\bibinfo{booktitle}{{Cargese Summer School on QCD Perspectives on Hot
  and Dense Matter}}} (\bibinfo{year}{2002}), pp. \bibinfo{pages}{73--145},
  \eprint{hep-ph/0202270}.

\bibitem[{\citenamefont{Jalilian-Marian and
  Kovchegov}(2006)}]{Jalilian-Marian:2005ccm}
\bibinfo{author}{\bibfnamefont{J.}~\bibnamefont{Jalilian-Marian}}
  \bibnamefont{and} \bibinfo{author}{\bibfnamefont{Y.~V.}
  \bibnamefont{Kovchegov}}, \bibinfo{journal}{Prog. Part. Nucl. Phys.}
  \textbf{\bibinfo{volume}{56}}, \bibinfo{pages}{104} (\bibinfo{year}{2006}),
  \eprint{hep-ph/0505052}.

\bibitem[{\citenamefont{Weigert}(2005)}]{Weigert:2005us}
\bibinfo{author}{\bibfnamefont{H.}~\bibnamefont{Weigert}},
  \bibinfo{journal}{Prog. Part. Nucl. Phys.} \textbf{\bibinfo{volume}{55}},
  \bibinfo{pages}{461} (\bibinfo{year}{2005}), \eprint{hep-ph/0501087}.

\bibitem[{\citenamefont{Morreale and Salazar}(2021)}]{Morreale:2021pnn}
\bibinfo{author}{\bibfnamefont{A.}~\bibnamefont{Morreale}} \bibnamefont{and}
  \bibinfo{author}{\bibfnamefont{F.}~\bibnamefont{Salazar}},
  \bibinfo{journal}{Universe} \textbf{\bibinfo{volume}{7}},
  \bibinfo{pages}{312} (\bibinfo{year}{2021}), \eprint{2108.08254}.

\bibitem[{\citenamefont{Kovner and Wiedemann}(2001)}]{Kovner:2001vi}
\bibinfo{author}{\bibfnamefont{A.}~\bibnamefont{Kovner}} \bibnamefont{and}
  \bibinfo{author}{\bibfnamefont{U.~A.} \bibnamefont{Wiedemann}},
  \bibinfo{journal}{Phys. Rev. D} \textbf{\bibinfo{volume}{64}},
  \bibinfo{pages}{114002} (\bibinfo{year}{2001}), \eprint{hep-ph/0106240}.

\bibitem[{\citenamefont{Jalilian-Marian and
  Kovchegov}(2004)}]{Jalilian-Marian:2004vhw}
\bibinfo{author}{\bibfnamefont{J.}~\bibnamefont{Jalilian-Marian}}
  \bibnamefont{and} \bibinfo{author}{\bibfnamefont{Y.~V.}
  \bibnamefont{Kovchegov}}, \bibinfo{journal}{Phys. Rev. D}
  \textbf{\bibinfo{volume}{70}}, \bibinfo{pages}{114017}
  (\bibinfo{year}{2004}), \bibinfo{note}{[Erratum: Phys.Rev.D 71, 079901
  (2005)]}, \eprint{hep-ph/0405266}.

\bibitem[{\citenamefont{Dumitru et~al.}(2006)\citenamefont{Dumitru,
  Hayashigaki, and Jalilian-Marian}}]{Dumitru:2005gt}
\bibinfo{author}{\bibfnamefont{A.}~\bibnamefont{Dumitru}},
  \bibinfo{author}{\bibfnamefont{A.}~\bibnamefont{Hayashigaki}},
  \bibnamefont{and}
  \bibinfo{author}{\bibfnamefont{J.}~\bibnamefont{Jalilian-Marian}},
  \bibinfo{journal}{Nucl. Phys. A} \textbf{\bibinfo{volume}{765}},
  \bibinfo{pages}{464} (\bibinfo{year}{2006}), \eprint{hep-ph/0506308}.

\bibitem[{\citenamefont{Jalilian-Marian}(2006)}]{Jalilian-Marian:2005qbq}
\bibinfo{author}{\bibfnamefont{J.}~\bibnamefont{Jalilian-Marian}},
  \bibinfo{journal}{Nucl. Phys. A} \textbf{\bibinfo{volume}{770}},
  \bibinfo{pages}{210} (\bibinfo{year}{2006}), \eprint{hep-ph/0509338}.

\bibitem[{\citenamefont{Marquet}(2007)}]{Marquet:2007vb}
\bibinfo{author}{\bibfnamefont{C.}~\bibnamefont{Marquet}},
  \bibinfo{journal}{Nucl. Phys. A} \textbf{\bibinfo{volume}{796}},
  \bibinfo{pages}{41} (\bibinfo{year}{2007}), \eprint{0708.0231}.

\bibitem[{\citenamefont{Albacete and Marquet}(2010)}]{Albacete:2010pg}
\bibinfo{author}{\bibfnamefont{J.~L.} \bibnamefont{Albacete}} \bibnamefont{and}
  \bibinfo{author}{\bibfnamefont{C.}~\bibnamefont{Marquet}},
  \bibinfo{journal}{Phys. Rev. Lett.} \textbf{\bibinfo{volume}{105}},
  \bibinfo{pages}{162301} (\bibinfo{year}{2010}), \eprint{1005.4065}.

\bibitem[{\citenamefont{Stasto et~al.}(2012)\citenamefont{Stasto, Xiao, and
  Yuan}}]{Stasto:2011ru}
\bibinfo{author}{\bibfnamefont{A.}~\bibnamefont{Stasto}},
  \bibinfo{author}{\bibfnamefont{B.-W.} \bibnamefont{Xiao}}, \bibnamefont{and}
  \bibinfo{author}{\bibfnamefont{F.}~\bibnamefont{Yuan}},
  \bibinfo{journal}{Phys. Lett. B} \textbf{\bibinfo{volume}{716}},
  \bibinfo{pages}{430} (\bibinfo{year}{2012}), \eprint{1109.1817}.

\bibitem[{\citenamefont{Lappi and Mantysaari}(2013)}]{Lappi:2012nh}
\bibinfo{author}{\bibfnamefont{T.}~\bibnamefont{Lappi}} \bibnamefont{and}
  \bibinfo{author}{\bibfnamefont{H.}~\bibnamefont{Mantysaari}},
  \bibinfo{journal}{Nucl. Phys. A} \textbf{\bibinfo{volume}{908}},
  \bibinfo{pages}{51} (\bibinfo{year}{2013}), \eprint{1209.2853}.

\bibitem[{\citenamefont{Jalilian-Marian and
  Rezaeian}(2012{\natexlab{a}})}]{Jalilian-Marian:2012wwi}
\bibinfo{author}{\bibfnamefont{J.}~\bibnamefont{Jalilian-Marian}}
  \bibnamefont{and} \bibinfo{author}{\bibfnamefont{A.~H.}
  \bibnamefont{Rezaeian}}, \bibinfo{journal}{Phys. Rev. D}
  \textbf{\bibinfo{volume}{86}}, \bibinfo{pages}{034016}
  (\bibinfo{year}{2012}{\natexlab{a}}), \eprint{1204.1319}.

\bibitem[{\citenamefont{Jalilian-Marian and
  Rezaeian}(2012{\natexlab{b}})}]{Jalilian-Marian:2011tvq}
\bibinfo{author}{\bibfnamefont{J.}~\bibnamefont{Jalilian-Marian}}
  \bibnamefont{and} \bibinfo{author}{\bibfnamefont{A.~H.}
  \bibnamefont{Rezaeian}}, \bibinfo{journal}{Phys. Rev. D}
  \textbf{\bibinfo{volume}{85}}, \bibinfo{pages}{014017}
  (\bibinfo{year}{2012}{\natexlab{b}}), \eprint{1110.2810}.

\bibitem[{\citenamefont{Zheng et~al.}(2014)\citenamefont{Zheng, Aschenauer,
  Lee, and Xiao}}]{Zheng:2014vka}
\bibinfo{author}{\bibfnamefont{L.}~\bibnamefont{Zheng}},
  \bibinfo{author}{\bibfnamefont{E.~C.} \bibnamefont{Aschenauer}},
  \bibinfo{author}{\bibfnamefont{J.~H.} \bibnamefont{Lee}}, \bibnamefont{and}
  \bibinfo{author}{\bibfnamefont{B.-W.} \bibnamefont{Xiao}},
  \bibinfo{journal}{Phys. Rev. D} \textbf{\bibinfo{volume}{89}},
  \bibinfo{pages}{074037} (\bibinfo{year}{2014}), \eprint{1403.2413}.

\bibitem[{\citenamefont{Stasto et~al.}(2018)\citenamefont{Stasto, Wei, Xiao,
  and Yuan}}]{Stasto:2018rci}
\bibinfo{author}{\bibfnamefont{A.}~\bibnamefont{Stasto}},
  \bibinfo{author}{\bibfnamefont{S.-Y.} \bibnamefont{Wei}},
  \bibinfo{author}{\bibfnamefont{B.-W.} \bibnamefont{Xiao}}, \bibnamefont{and}
  \bibinfo{author}{\bibfnamefont{F.}~\bibnamefont{Yuan}},
  \bibinfo{journal}{Phys. Lett. B} \textbf{\bibinfo{volume}{784}},
  \bibinfo{pages}{301} (\bibinfo{year}{2018}), \eprint{1805.05712}.

\bibitem[{\citenamefont{Albacete et~al.}(2019)\citenamefont{Albacete,
  Giacalone, Marquet, and Matas}}]{Albacete:2018ruq}
\bibinfo{author}{\bibfnamefont{J.~L.} \bibnamefont{Albacete}},
  \bibinfo{author}{\bibfnamefont{G.}~\bibnamefont{Giacalone}},
  \bibinfo{author}{\bibfnamefont{C.}~\bibnamefont{Marquet}}, \bibnamefont{and}
  \bibinfo{author}{\bibfnamefont{M.}~\bibnamefont{Matas}},
  \bibinfo{journal}{Phys. Rev. D} \textbf{\bibinfo{volume}{99}},
  \bibinfo{pages}{014002} (\bibinfo{year}{2019}), \eprint{1805.05711}.

\bibitem[{\citenamefont{M\"antysaari et~al.}(2020)\citenamefont{M\"antysaari,
  Mueller, Salazar, and Schenke}}]{Mantysaari:2019hkq}
\bibinfo{author}{\bibfnamefont{H.}~\bibnamefont{M\"antysaari}},
  \bibinfo{author}{\bibfnamefont{N.}~\bibnamefont{Mueller}},
  \bibinfo{author}{\bibfnamefont{F.}~\bibnamefont{Salazar}}, \bibnamefont{and}
  \bibinfo{author}{\bibfnamefont{B.}~\bibnamefont{Schenke}},
  \bibinfo{journal}{Phys. Rev. Lett.} \textbf{\bibinfo{volume}{124}},
  \bibinfo{pages}{112301} (\bibinfo{year}{2020}), \eprint{1912.05586}.

\bibitem[{\citenamefont{Hatta et~al.}(2021{\natexlab{a}})\citenamefont{Hatta,
  Xiao, Yuan, and Zhou}}]{Hatta:2020bgy}
\bibinfo{author}{\bibfnamefont{Y.}~\bibnamefont{Hatta}},
  \bibinfo{author}{\bibfnamefont{B.-W.} \bibnamefont{Xiao}},
  \bibinfo{author}{\bibfnamefont{F.}~\bibnamefont{Yuan}}, \bibnamefont{and}
  \bibinfo{author}{\bibfnamefont{J.}~\bibnamefont{Zhou}},
  \bibinfo{journal}{Phys. Rev. Lett.} \textbf{\bibinfo{volume}{126}},
  \bibinfo{pages}{142001} (\bibinfo{year}{2021}{\natexlab{a}}),
  \eprint{2010.10774}.

\bibitem[{\citenamefont{Jia et~al.}(2020)\citenamefont{Jia, Wei, Xiao, and
  Yuan}}]{Jia:2019qbl}
\bibinfo{author}{\bibfnamefont{J.}~\bibnamefont{Jia}},
  \bibinfo{author}{\bibfnamefont{S.-Y.} \bibnamefont{Wei}},
  \bibinfo{author}{\bibfnamefont{B.-W.} \bibnamefont{Xiao}}, \bibnamefont{and}
  \bibinfo{author}{\bibfnamefont{F.}~\bibnamefont{Yuan}},
  \bibinfo{journal}{Phys. Rev. D} \textbf{\bibinfo{volume}{101}},
  \bibinfo{pages}{094008} (\bibinfo{year}{2020}), \eprint{1910.05290}.

\bibitem[{\citenamefont{Gelis and Jalilian-Marian}(2002)}]{Gelis:2002fw}
\bibinfo{author}{\bibfnamefont{F.}~\bibnamefont{Gelis}} \bibnamefont{and}
  \bibinfo{author}{\bibfnamefont{J.}~\bibnamefont{Jalilian-Marian}},
  \bibinfo{journal}{Phys. Rev. D} \textbf{\bibinfo{volume}{66}},
  \bibinfo{pages}{094014} (\bibinfo{year}{2002}), \eprint{hep-ph/0208141}.

\bibitem[{\citenamefont{Dominguez et~al.}(2011)\citenamefont{Dominguez,
  Marquet, Xiao, and Yuan}}]{Dominguez:2011wm}
\bibinfo{author}{\bibfnamefont{F.}~\bibnamefont{Dominguez}},
  \bibinfo{author}{\bibfnamefont{C.}~\bibnamefont{Marquet}},
  \bibinfo{author}{\bibfnamefont{B.-W.} \bibnamefont{Xiao}}, \bibnamefont{and}
  \bibinfo{author}{\bibfnamefont{F.}~\bibnamefont{Yuan}},
  \bibinfo{journal}{Phys. Rev. D} \textbf{\bibinfo{volume}{83}},
  \bibinfo{pages}{105005} (\bibinfo{year}{2011}), \eprint{1101.0715}.

\bibitem[{\citenamefont{Metz and Zhou}(2011)}]{Metz:2011wb}
\bibinfo{author}{\bibfnamefont{A.}~\bibnamefont{Metz}} \bibnamefont{and}
  \bibinfo{author}{\bibfnamefont{J.}~\bibnamefont{Zhou}},
  \bibinfo{journal}{Phys. Rev. D} \textbf{\bibinfo{volume}{84}},
  \bibinfo{pages}{051503} (\bibinfo{year}{2011}), \eprint{1105.1991}.

\bibitem[{\citenamefont{Dominguez et~al.}(2012)\citenamefont{Dominguez, Qiu,
  Xiao, and Yuan}}]{Dominguez:2011br}
\bibinfo{author}{\bibfnamefont{F.}~\bibnamefont{Dominguez}},
  \bibinfo{author}{\bibfnamefont{J.-W.} \bibnamefont{Qiu}},
  \bibinfo{author}{\bibfnamefont{B.-W.} \bibnamefont{Xiao}}, \bibnamefont{and}
  \bibinfo{author}{\bibfnamefont{F.}~\bibnamefont{Yuan}},
  \bibinfo{journal}{Phys. Rev. D} \textbf{\bibinfo{volume}{85}},
  \bibinfo{pages}{045003} (\bibinfo{year}{2012}), \eprint{1109.6293}.

\bibitem[{\citenamefont{Iancu and Laidet}(2013)}]{Iancu:2013dta}
\bibinfo{author}{\bibfnamefont{E.}~\bibnamefont{Iancu}} \bibnamefont{and}
  \bibinfo{author}{\bibfnamefont{J.}~\bibnamefont{Laidet}},
  \bibinfo{journal}{Nucl. Phys. A} \textbf{\bibinfo{volume}{916}},
  \bibinfo{pages}{48} (\bibinfo{year}{2013}), \eprint{1305.5926}.

\bibitem[{\citenamefont{Altinoluk
  et~al.}(2016{\natexlab{a}})\citenamefont{Altinoluk, Armesto, Beuf, and
  Rezaeian}}]{Altinoluk:2015dpi}
\bibinfo{author}{\bibfnamefont{T.}~\bibnamefont{Altinoluk}},
  \bibinfo{author}{\bibfnamefont{N.}~\bibnamefont{Armesto}},
  \bibinfo{author}{\bibfnamefont{G.}~\bibnamefont{Beuf}}, \bibnamefont{and}
  \bibinfo{author}{\bibfnamefont{A.~H.} \bibnamefont{Rezaeian}},
  \bibinfo{journal}{Phys. Lett. B} \textbf{\bibinfo{volume}{758}},
  \bibinfo{pages}{373} (\bibinfo{year}{2016}{\natexlab{a}}),
  \eprint{1511.07452}.

\bibitem[{\citenamefont{Hatta et~al.}(2016)\citenamefont{Hatta, Xiao, and
  Yuan}}]{Hatta:2016dxp}
\bibinfo{author}{\bibfnamefont{Y.}~\bibnamefont{Hatta}},
  \bibinfo{author}{\bibfnamefont{B.-W.} \bibnamefont{Xiao}}, \bibnamefont{and}
  \bibinfo{author}{\bibfnamefont{F.}~\bibnamefont{Yuan}},
  \bibinfo{journal}{Phys. Rev. Lett.} \textbf{\bibinfo{volume}{116}},
  \bibinfo{pages}{202301} (\bibinfo{year}{2016}), \eprint{1601.01585}.

\bibitem[{\citenamefont{Dumitru et~al.}(2015)\citenamefont{Dumitru, Lappi, and
  Skokov}}]{Dumitru:2015gaa}
\bibinfo{author}{\bibfnamefont{A.}~\bibnamefont{Dumitru}},
  \bibinfo{author}{\bibfnamefont{T.}~\bibnamefont{Lappi}}, \bibnamefont{and}
  \bibinfo{author}{\bibfnamefont{V.}~\bibnamefont{Skokov}},
  \bibinfo{journal}{Phys. Rev. Lett.} \textbf{\bibinfo{volume}{115}},
  \bibinfo{pages}{252301} (\bibinfo{year}{2015}), \eprint{1508.04438}.

\bibitem[{\citenamefont{Kotko et~al.}(2015)\citenamefont{Kotko, Kutak, Marquet,
  Petreska, Sapeta, and van Hameren}}]{Kotko:2015ura}
\bibinfo{author}{\bibfnamefont{P.}~\bibnamefont{Kotko}},
  \bibinfo{author}{\bibfnamefont{K.}~\bibnamefont{Kutak}},
  \bibinfo{author}{\bibfnamefont{C.}~\bibnamefont{Marquet}},
  \bibinfo{author}{\bibfnamefont{E.}~\bibnamefont{Petreska}},
  \bibinfo{author}{\bibfnamefont{S.}~\bibnamefont{Sapeta}}, \bibnamefont{and}
  \bibinfo{author}{\bibfnamefont{A.}~\bibnamefont{van Hameren}},
  \bibinfo{journal}{JHEP} \textbf{\bibinfo{volume}{09}}, \bibinfo{pages}{106}
  (\bibinfo{year}{2015}), \eprint{1503.03421}.

\bibitem[{\citenamefont{Marquet et~al.}(2016)\citenamefont{Marquet, Petreska,
  and Roiesnel}}]{Marquet:2016cgx}
\bibinfo{author}{\bibfnamefont{C.}~\bibnamefont{Marquet}},
  \bibinfo{author}{\bibfnamefont{E.}~\bibnamefont{Petreska}}, \bibnamefont{and}
  \bibinfo{author}{\bibfnamefont{C.}~\bibnamefont{Roiesnel}},
  \bibinfo{journal}{JHEP} \textbf{\bibinfo{volume}{10}}, \bibinfo{pages}{065}
  (\bibinfo{year}{2016}), \eprint{1608.02577}.

\bibitem[{\citenamefont{van Hameren et~al.}(2016)\citenamefont{van Hameren,
  Kotko, Kutak, Marquet, Petreska, and Sapeta}}]{vanHameren:2016ftb}
\bibinfo{author}{\bibfnamefont{A.}~\bibnamefont{van Hameren}},
  \bibinfo{author}{\bibfnamefont{P.}~\bibnamefont{Kotko}},
  \bibinfo{author}{\bibfnamefont{K.}~\bibnamefont{Kutak}},
  \bibinfo{author}{\bibfnamefont{C.}~\bibnamefont{Marquet}},
  \bibinfo{author}{\bibfnamefont{E.}~\bibnamefont{Petreska}}, \bibnamefont{and}
  \bibinfo{author}{\bibfnamefont{S.}~\bibnamefont{Sapeta}},
  \bibinfo{journal}{JHEP} \textbf{\bibinfo{volume}{12}}, \bibinfo{pages}{034}
  (\bibinfo{year}{2016}), \bibinfo{note}{[Erratum: JHEP 02, 158 (2019)]},
  \eprint{1607.03121}.

\bibitem[{\citenamefont{Marquet et~al.}(2018)\citenamefont{Marquet, Roiesnel,
  and Taels}}]{Marquet:2017xwy}
\bibinfo{author}{\bibfnamefont{C.}~\bibnamefont{Marquet}},
  \bibinfo{author}{\bibfnamefont{C.}~\bibnamefont{Roiesnel}}, \bibnamefont{and}
  \bibinfo{author}{\bibfnamefont{P.}~\bibnamefont{Taels}},
  \bibinfo{journal}{Phys. Rev. D} \textbf{\bibinfo{volume}{97}},
  \bibinfo{pages}{014004} (\bibinfo{year}{2018}), \eprint{1710.05698}.

\bibitem[{\citenamefont{Dumitru et~al.}(2019)\citenamefont{Dumitru, Skokov, and
  Ullrich}}]{Dumitru:2018kuw}
\bibinfo{author}{\bibfnamefont{A.}~\bibnamefont{Dumitru}},
  \bibinfo{author}{\bibfnamefont{V.}~\bibnamefont{Skokov}}, \bibnamefont{and}
  \bibinfo{author}{\bibfnamefont{T.}~\bibnamefont{Ullrich}},
  \bibinfo{journal}{Phys. Rev. C} \textbf{\bibinfo{volume}{99}},
  \bibinfo{pages}{015204} (\bibinfo{year}{2019}), \eprint{1809.02615}.

\bibitem[{\citenamefont{Dumitru and
  Jalilian-Marian}(2002{\natexlab{a}})}]{Dumitru:2001jn}
\bibinfo{author}{\bibfnamefont{A.}~\bibnamefont{Dumitru}} \bibnamefont{and}
  \bibinfo{author}{\bibfnamefont{J.}~\bibnamefont{Jalilian-Marian}},
  \bibinfo{journal}{Phys. Lett. B} \textbf{\bibinfo{volume}{547}},
  \bibinfo{pages}{15} (\bibinfo{year}{2002}{\natexlab{a}}),
  \eprint{hep-ph/0111357}.

\bibitem[{\citenamefont{Dumitru and
  Jalilian-Marian}(2002{\natexlab{b}})}]{Dumitru:2002qt}
\bibinfo{author}{\bibfnamefont{A.}~\bibnamefont{Dumitru}} \bibnamefont{and}
  \bibinfo{author}{\bibfnamefont{J.}~\bibnamefont{Jalilian-Marian}},
  \bibinfo{journal}{Phys. Rev. Lett.} \textbf{\bibinfo{volume}{89}},
  \bibinfo{pages}{022301} (\bibinfo{year}{2002}{\natexlab{b}}),
  \eprint{hep-ph/0204028}.

\bibitem[{\citenamefont{M\"antysaari et~al.}(2019)\citenamefont{M\"antysaari,
  Mueller, and Schenke}}]{Mantysaari:2019csc}
\bibinfo{author}{\bibfnamefont{H.}~\bibnamefont{M\"antysaari}},
  \bibinfo{author}{\bibfnamefont{N.}~\bibnamefont{Mueller}}, \bibnamefont{and}
  \bibinfo{author}{\bibfnamefont{B.}~\bibnamefont{Schenke}},
  \bibinfo{journal}{Phys. Rev. D} \textbf{\bibinfo{volume}{99}},
  \bibinfo{pages}{074004} (\bibinfo{year}{2019}), \eprint{1902.05087}.

\bibitem[{\citenamefont{Salazar and Schenke}(2019)}]{Salazar:2019ncp}
\bibinfo{author}{\bibfnamefont{F.}~\bibnamefont{Salazar}} \bibnamefont{and}
  \bibinfo{author}{\bibfnamefont{B.}~\bibnamefont{Schenke}},
  \bibinfo{journal}{Phys. Rev. D} \textbf{\bibinfo{volume}{100}},
  \bibinfo{pages}{034007} (\bibinfo{year}{2019}), \eprint{1905.03763}.

\bibitem[{\citenamefont{Boussarie et~al.}(2021)\citenamefont{Boussarie,
  M\"antysaari, Salazar, and Schenke}}]{Boussarie:2021ybe}
\bibinfo{author}{\bibfnamefont{R.}~\bibnamefont{Boussarie}},
  \bibinfo{author}{\bibfnamefont{H.}~\bibnamefont{M\"antysaari}},
  \bibinfo{author}{\bibfnamefont{F.}~\bibnamefont{Salazar}}, \bibnamefont{and}
  \bibinfo{author}{\bibfnamefont{B.}~\bibnamefont{Schenke}},
  \bibinfo{journal}{JHEP} \textbf{\bibinfo{volume}{09}}, \bibinfo{pages}{178}
  (\bibinfo{year}{2021}), \eprint{2106.11301}.

\bibitem[{\citenamefont{Ayala et~al.}(1996)\citenamefont{Ayala,
  Jalilian-Marian, McLerran, and Venugopalan}}]{Ayala:1995hx}
\bibinfo{author}{\bibfnamefont{A.}~\bibnamefont{Ayala}},
  \bibinfo{author}{\bibfnamefont{J.}~\bibnamefont{Jalilian-Marian}},
  \bibinfo{author}{\bibfnamefont{L.~D.} \bibnamefont{McLerran}},
  \bibnamefont{and}
  \bibinfo{author}{\bibfnamefont{R.}~\bibnamefont{Venugopalan}},
  \bibinfo{journal}{Phys. Rev. D} \textbf{\bibinfo{volume}{53}},
  \bibinfo{pages}{458} (\bibinfo{year}{1996}), \eprint{hep-ph/9508302}.

\bibitem[{\citenamefont{Jalilian-Marian
  et~al.}(1997{\natexlab{a}})\citenamefont{Jalilian-Marian, Kovner, McLerran,
  and Weigert}}]{Jalilian-Marian:1996mkd}
\bibinfo{author}{\bibfnamefont{J.}~\bibnamefont{Jalilian-Marian}},
  \bibinfo{author}{\bibfnamefont{A.}~\bibnamefont{Kovner}},
  \bibinfo{author}{\bibfnamefont{L.~D.} \bibnamefont{McLerran}},
  \bibnamefont{and} \bibinfo{author}{\bibfnamefont{H.}~\bibnamefont{Weigert}},
  \bibinfo{journal}{Phys. Rev. D} \textbf{\bibinfo{volume}{55}},
  \bibinfo{pages}{5414} (\bibinfo{year}{1997}{\natexlab{a}}),
  \eprint{hep-ph/9606337}.

\bibitem[{\citenamefont{Kotko et~al.}(2017)\citenamefont{Kotko, Kutak, Sapeta,
  Stasto, and Strikman}}]{Kotko:2017oxg}
\bibinfo{author}{\bibfnamefont{P.}~\bibnamefont{Kotko}},
  \bibinfo{author}{\bibfnamefont{K.}~\bibnamefont{Kutak}},
  \bibinfo{author}{\bibfnamefont{S.}~\bibnamefont{Sapeta}},
  \bibinfo{author}{\bibfnamefont{A.~M.} \bibnamefont{Stasto}},
  \bibnamefont{and} \bibinfo{author}{\bibfnamefont{M.}~\bibnamefont{Strikman}},
  \bibinfo{journal}{Eur. Phys. J. C} \textbf{\bibinfo{volume}{77}},
  \bibinfo{pages}{353} (\bibinfo{year}{2017}), \eprint{1702.03063}.

\bibitem[{\citenamefont{Hagiwara et~al.}(2017)\citenamefont{Hagiwara, Hatta,
  Pasechnik, Tasevsky, and Teryaev}}]{Hagiwara:2017fye}
\bibinfo{author}{\bibfnamefont{Y.}~\bibnamefont{Hagiwara}},
  \bibinfo{author}{\bibfnamefont{Y.}~\bibnamefont{Hatta}},
  \bibinfo{author}{\bibfnamefont{R.}~\bibnamefont{Pasechnik}},
  \bibinfo{author}{\bibfnamefont{M.}~\bibnamefont{Tasevsky}}, \bibnamefont{and}
  \bibinfo{author}{\bibfnamefont{O.}~\bibnamefont{Teryaev}},
  \bibinfo{journal}{Phys. Rev. D} \textbf{\bibinfo{volume}{96}},
  \bibinfo{pages}{034009} (\bibinfo{year}{2017}), \eprint{1706.01765}.

\bibitem[{\citenamefont{Henley and Jalilian-Marian}(2006)}]{Henley:2005ms}
\bibinfo{author}{\bibfnamefont{E.~M.} \bibnamefont{Henley}} \bibnamefont{and}
  \bibinfo{author}{\bibfnamefont{J.}~\bibnamefont{Jalilian-Marian}},
  \bibinfo{journal}{Phys. Rev. D} \textbf{\bibinfo{volume}{73}},
  \bibinfo{pages}{094004} (\bibinfo{year}{2006}), \eprint{hep-ph/0512220}.

\bibitem[{\citenamefont{Klein and M\"antysaari}(2019)}]{Klein:2019qfb}
\bibinfo{author}{\bibfnamefont{S.~R.} \bibnamefont{Klein}} \bibnamefont{and}
  \bibinfo{author}{\bibfnamefont{H.}~\bibnamefont{M\"antysaari}},
  \bibinfo{journal}{Nature Rev. Phys.} \textbf{\bibinfo{volume}{1}},
  \bibinfo{pages}{662} (\bibinfo{year}{2019}), \eprint{1910.10858}.

\bibitem[{\citenamefont{Hatta et~al.}(2021{\natexlab{b}})\citenamefont{Hatta,
  Xiao, Yuan, and Zhou}}]{Hatta:2021jcd}
\bibinfo{author}{\bibfnamefont{Y.}~\bibnamefont{Hatta}},
  \bibinfo{author}{\bibfnamefont{B.-W.} \bibnamefont{Xiao}},
  \bibinfo{author}{\bibfnamefont{F.}~\bibnamefont{Yuan}}, \bibnamefont{and}
  \bibinfo{author}{\bibfnamefont{J.}~\bibnamefont{Zhou}},
  \bibinfo{journal}{Phys. Rev. D} \textbf{\bibinfo{volume}{104}},
  \bibinfo{pages}{054037} (\bibinfo{year}{2021}{\natexlab{b}}),
  \eprint{2106.05307}.

\bibitem[{\citenamefont{Kolb\'e et~al.}(2021)\citenamefont{Kolb\'e, Roy,
  Salazar, Schenke, and Venugopalan}}]{Kolbe:2020tlq}
\bibinfo{author}{\bibfnamefont{I.}~\bibnamefont{Kolb\'e}},
  \bibinfo{author}{\bibfnamefont{K.}~\bibnamefont{Roy}},
  \bibinfo{author}{\bibfnamefont{F.}~\bibnamefont{Salazar}},
  \bibinfo{author}{\bibfnamefont{B.}~\bibnamefont{Schenke}}, \bibnamefont{and}
  \bibinfo{author}{\bibfnamefont{R.}~\bibnamefont{Venugopalan}},
  \bibinfo{journal}{JHEP} \textbf{\bibinfo{volume}{01}}, \bibinfo{pages}{052}
  (\bibinfo{year}{2021}), \eprint{2008.04372}.

\bibitem[{\citenamefont{Altinoluk et~al.}(2019)\citenamefont{Altinoluk,
  Boussarie, and Kotko}}]{Altinoluk:2019fui}
\bibinfo{author}{\bibfnamefont{T.}~\bibnamefont{Altinoluk}},
  \bibinfo{author}{\bibfnamefont{R.}~\bibnamefont{Boussarie}},
  \bibnamefont{and} \bibinfo{author}{\bibfnamefont{P.}~\bibnamefont{Kotko}},
  \bibinfo{journal}{JHEP} \textbf{\bibinfo{volume}{05}}, \bibinfo{pages}{156}
  (\bibinfo{year}{2019}), \eprint{1901.01175}.

\bibitem[{\citenamefont{Boussarie et~al.}(2019)\citenamefont{Boussarie,
  Grabovsky, Szymanowski, and Wallon}}]{Boussarie:2019ero}
\bibinfo{author}{\bibfnamefont{R.}~\bibnamefont{Boussarie}},
  \bibinfo{author}{\bibfnamefont{A.~V.} \bibnamefont{Grabovsky}},
  \bibinfo{author}{\bibfnamefont{L.}~\bibnamefont{Szymanowski}},
  \bibnamefont{and} \bibinfo{author}{\bibfnamefont{S.}~\bibnamefont{Wallon}},
  \bibinfo{journal}{Phys. Rev. D} \textbf{\bibinfo{volume}{100}},
  \bibinfo{pages}{074020} (\bibinfo{year}{2019}), \eprint{1905.07371}.

\bibitem[{\citenamefont{Boussarie et~al.}(2016)\citenamefont{Boussarie,
  Grabovsky, Szymanowski, and Wallon}}]{Boussarie:2016ogo}
\bibinfo{author}{\bibfnamefont{R.}~\bibnamefont{Boussarie}},
  \bibinfo{author}{\bibfnamefont{A.~V.} \bibnamefont{Grabovsky}},
  \bibinfo{author}{\bibfnamefont{L.}~\bibnamefont{Szymanowski}},
  \bibnamefont{and} \bibinfo{author}{\bibfnamefont{S.}~\bibnamefont{Wallon}},
  \bibinfo{journal}{JHEP} \textbf{\bibinfo{volume}{11}}, \bibinfo{pages}{149}
  (\bibinfo{year}{2016}), \eprint{1606.00419}.

\bibitem[{\citenamefont{Boussarie et~al.}(2014)\citenamefont{Boussarie,
  Grabovsky, Szymanowski, and Wallon}}]{Boussarie:2014lxa}
\bibinfo{author}{\bibfnamefont{R.}~\bibnamefont{Boussarie}},
  \bibinfo{author}{\bibfnamefont{A.~V.} \bibnamefont{Grabovsky}},
  \bibinfo{author}{\bibfnamefont{L.}~\bibnamefont{Szymanowski}},
  \bibnamefont{and} \bibinfo{author}{\bibfnamefont{S.}~\bibnamefont{Wallon}},
  \bibinfo{journal}{JHEP} \textbf{\bibinfo{volume}{09}}, \bibinfo{pages}{026}
  (\bibinfo{year}{2014}), \eprint{1405.7676}.

\bibitem[{\citenamefont{Dumitru and Jalilian-Marian}(2010)}]{Dumitru:2010ak}
\bibinfo{author}{\bibfnamefont{A.}~\bibnamefont{Dumitru}} \bibnamefont{and}
  \bibinfo{author}{\bibfnamefont{J.}~\bibnamefont{Jalilian-Marian}},
  \bibinfo{journal}{Phys. Rev. D} \textbf{\bibinfo{volume}{82}},
  \bibinfo{pages}{074023} (\bibinfo{year}{2010}), \eprint{1008.0480}.

\bibitem[{\citenamefont{Fadin and Lipatov}(1998)}]{Fadin:1998py}
\bibinfo{author}{\bibfnamefont{V.~S.} \bibnamefont{Fadin}} \bibnamefont{and}
  \bibinfo{author}{\bibfnamefont{L.~N.} \bibnamefont{Lipatov}},
  \bibinfo{journal}{Phys. Lett. B} \textbf{\bibinfo{volume}{429}},
  \bibinfo{pages}{127} (\bibinfo{year}{1998}), \eprint{hep-ph/9802290}.

\bibitem[{\citenamefont{Chirilli
  et~al.}(2012{\natexlab{a}})\citenamefont{Chirilli, Xiao, and
  Yuan}}]{Chirilli:2011km}
\bibinfo{author}{\bibfnamefont{G.~A.} \bibnamefont{Chirilli}},
  \bibinfo{author}{\bibfnamefont{B.-W.} \bibnamefont{Xiao}}, \bibnamefont{and}
  \bibinfo{author}{\bibfnamefont{F.}~\bibnamefont{Yuan}},
  \bibinfo{journal}{Phys. Rev. Lett.} \textbf{\bibinfo{volume}{108}},
  \bibinfo{pages}{122301} (\bibinfo{year}{2012}{\natexlab{a}}),
  \eprint{1112.1061}.

\bibitem[{\citenamefont{Chirilli
  et~al.}(2012{\natexlab{b}})\citenamefont{Chirilli, Xiao, and
  Yuan}}]{Chirilli:2012jd}
\bibinfo{author}{\bibfnamefont{G.~A.} \bibnamefont{Chirilli}},
  \bibinfo{author}{\bibfnamefont{B.-W.} \bibnamefont{Xiao}}, \bibnamefont{and}
  \bibinfo{author}{\bibfnamefont{F.}~\bibnamefont{Yuan}},
  \bibinfo{journal}{Phys. Rev. D} \textbf{\bibinfo{volume}{86}},
  \bibinfo{pages}{054005} (\bibinfo{year}{2012}{\natexlab{b}}),
  \eprint{1203.6139}.

\bibitem[{\citenamefont{Balitsky and
  Chirilli}(2013{\natexlab{a}})}]{balitsky:2012bs}
\bibinfo{author}{\bibfnamefont{I.}~\bibnamefont{Balitsky}} \bibnamefont{and}
  \bibinfo{author}{\bibfnamefont{G.~A.} \bibnamefont{Chirilli}},
  \bibinfo{journal}{Phys. Rev. D} \textbf{\bibinfo{volume}{87}},
  \bibinfo{pages}{014013} (\bibinfo{year}{2013}{\natexlab{a}}),
  \eprint{1207.3844}.

\bibitem[{\citenamefont{Balitsky and
  Chirilli}(2013{\natexlab{b}})}]{Balitsky:2013fea}
\bibinfo{author}{\bibfnamefont{I.}~\bibnamefont{Balitsky}} \bibnamefont{and}
  \bibinfo{author}{\bibfnamefont{G.~A.} \bibnamefont{Chirilli}},
  \bibinfo{journal}{Phys. Rev. D} \textbf{\bibinfo{volume}{88}},
  \bibinfo{pages}{111501} (\bibinfo{year}{2013}{\natexlab{b}}),
  \eprint{1309.7644}.

\bibitem[{\citenamefont{Grabovsky}(2013)}]{grabovsky:2013mba}
\bibinfo{author}{\bibfnamefont{A.~V.} \bibnamefont{Grabovsky}},
  \bibinfo{journal}{JHEP} \textbf{\bibinfo{volume}{09}}, \bibinfo{pages}{141}
  (\bibinfo{year}{2013}), \eprint{1307.5414}.

\bibitem[{\citenamefont{Caron-Huot}(2015)}]{caron-huot:2013fea}
\bibinfo{author}{\bibfnamefont{S.}~\bibnamefont{Caron-Huot}},
  \bibinfo{journal}{JHEP} \textbf{\bibinfo{volume}{05}}, \bibinfo{pages}{093}
  (\bibinfo{year}{2015}), \eprint{1309.6521}.

\bibitem[{\citenamefont{Kovner et~al.}(2014)\citenamefont{Kovner, Lublinsky,
  and Mulian}}]{kovner:2013ona}
\bibinfo{author}{\bibfnamefont{A.}~\bibnamefont{Kovner}},
  \bibinfo{author}{\bibfnamefont{M.}~\bibnamefont{Lublinsky}},
  \bibnamefont{and} \bibinfo{author}{\bibfnamefont{Y.}~\bibnamefont{Mulian}},
  \bibinfo{journal}{Phys. Rev. D} \textbf{\bibinfo{volume}{89}},
  \bibinfo{pages}{061704} (\bibinfo{year}{2014}), \eprint{1310.0378}.

\bibitem[{\citenamefont{Lublinsky and Mulian}(2017)}]{lublinsky:2016meo}
\bibinfo{author}{\bibfnamefont{M.}~\bibnamefont{Lublinsky}} \bibnamefont{and}
  \bibinfo{author}{\bibfnamefont{Y.}~\bibnamefont{Mulian}},
  \bibinfo{journal}{JHEP} \textbf{\bibinfo{volume}{05}}, \bibinfo{pages}{097}
  (\bibinfo{year}{2017}), \eprint{1610.03453}.

\bibitem[{\citenamefont{Caron-Huot and Herranen}(2018)}]{caron-huot:2016tzz}
\bibinfo{author}{\bibfnamefont{S.}~\bibnamefont{Caron-Huot}} \bibnamefont{and}
  \bibinfo{author}{\bibfnamefont{M.}~\bibnamefont{Herranen}},
  \bibinfo{journal}{JHEP} \textbf{\bibinfo{volume}{02}}, \bibinfo{pages}{058}
  (\bibinfo{year}{2018}), \eprint{1604.07417}.

\bibitem[{\citenamefont{Boussarie et~al.}(2018)\citenamefont{Boussarie,
  Grabovsky, Ivanov, Szymanowski, and Wallon}}]{boussarie:2017dmx}
\bibinfo{author}{\bibfnamefont{R.}~\bibnamefont{Boussarie}},
  \bibinfo{author}{\bibfnamefont{A.~V.} \bibnamefont{Grabovsky}},
  \bibinfo{author}{\bibfnamefont{D.~Y.} \bibnamefont{Ivanov}},
  \bibinfo{author}{\bibfnamefont{L.}~\bibnamefont{Szymanowski}},
  \bibnamefont{and} \bibinfo{author}{\bibfnamefont{S.}~\bibnamefont{Wallon}},
  \bibinfo{journal}{PoS} \textbf{\bibinfo{volume}{DIS2017}},
  \bibinfo{pages}{062} (\bibinfo{year}{2018}), \eprint{1709.04422}.

\bibitem[{\citenamefont{Beuf et~al.}(2022{\natexlab{a}})\citenamefont{Beuf,
  Lappi, and Paatelainen}}]{beuf:2022ndu}
\bibinfo{author}{\bibfnamefont{G.}~\bibnamefont{Beuf}},
  \bibinfo{author}{\bibfnamefont{T.}~\bibnamefont{Lappi}}, \bibnamefont{and}
  \bibinfo{author}{\bibfnamefont{R.}~\bibnamefont{Paatelainen}},
  \bibinfo{journal}{Phys. Rev. D} \textbf{\bibinfo{volume}{106}},
  \bibinfo{pages}{034013} (\bibinfo{year}{2022}{\natexlab{a}}),
  \eprint{2204.02486}.

\bibitem[{\citenamefont{Beuf et~al.}(2022{\natexlab{b}})\citenamefont{Beuf,
  Lappi, and Paatelainen}}]{beuf:2021srj}
\bibinfo{author}{\bibfnamefont{G.}~\bibnamefont{Beuf}},
  \bibinfo{author}{\bibfnamefont{T.}~\bibnamefont{Lappi}}, \bibnamefont{and}
  \bibinfo{author}{\bibfnamefont{R.}~\bibnamefont{Paatelainen}},
  \bibinfo{journal}{Phys. Rev. Lett.} \textbf{\bibinfo{volume}{129}},
  \bibinfo{pages}{072001} (\bibinfo{year}{2022}{\natexlab{b}}),
  \eprint{2112.03158}.

\bibitem[{\citenamefont{Beuf et~al.}(2021)\citenamefont{Beuf, Lappi, and
  Paatelainen}}]{beuf:2021qqa}
\bibinfo{author}{\bibfnamefont{G.}~\bibnamefont{Beuf}},
  \bibinfo{author}{\bibfnamefont{T.}~\bibnamefont{Lappi}}, \bibnamefont{and}
  \bibinfo{author}{\bibfnamefont{R.}~\bibnamefont{Paatelainen}},
  \bibinfo{journal}{Phys. Rev. D} \textbf{\bibinfo{volume}{104}},
  \bibinfo{pages}{056032} (\bibinfo{year}{2021}), \eprint{2103.14549}.

\bibitem[{\citenamefont{M\"antysaari and Penttala}(2021)}]{mantysaari:2021ryb}
\bibinfo{author}{\bibfnamefont{H.}~\bibnamefont{M\"antysaari}}
  \bibnamefont{and} \bibinfo{author}{\bibfnamefont{J.}~\bibnamefont{Penttala}},
  \bibinfo{journal}{Phys. Lett. B} \textbf{\bibinfo{volume}{823}},
  \bibinfo{pages}{136723} (\bibinfo{year}{2021}), \eprint{2104.02349}.

\bibitem[{\citenamefont{M\"antysaari and
  Penttala}(2022{\natexlab{a}})}]{mantysaari:2022bsp}
\bibinfo{author}{\bibfnamefont{H.}~\bibnamefont{M\"antysaari}}
  \bibnamefont{and} \bibinfo{author}{\bibfnamefont{J.}~\bibnamefont{Penttala}},
  \bibinfo{journal}{Phys. Rev. D} \textbf{\bibinfo{volume}{105}},
  \bibinfo{pages}{114038} (\bibinfo{year}{2022}{\natexlab{a}}),
  \eprint{2203.16911}.

\bibitem[{\citenamefont{M\"antysaari and
  Penttala}(2022{\natexlab{b}})}]{mantysaari:2022kdm}
\bibinfo{author}{\bibfnamefont{H.}~\bibnamefont{M\"antysaari}}
  \bibnamefont{and} \bibinfo{author}{\bibfnamefont{J.}~\bibnamefont{Penttala}},
  \bibinfo{journal}{JHEP} \textbf{\bibinfo{volume}{08}}, \bibinfo{pages}{247}
  (\bibinfo{year}{2022}{\natexlab{b}}), \eprint{2204.14031}.

\bibitem[{\citenamefont{Lappi et~al.}(2022)\citenamefont{Lappi, M\"antysaari,
  and Penttala}}]{lappi:2021oag}
\bibinfo{author}{\bibfnamefont{T.}~\bibnamefont{Lappi}},
  \bibinfo{author}{\bibfnamefont{H.}~\bibnamefont{M\"antysaari}},
  \bibnamefont{and} \bibinfo{author}{\bibfnamefont{J.}~\bibnamefont{Penttala}},
  \bibinfo{journal}{SciPost Phys. Proc.} \textbf{\bibinfo{volume}{8}},
  \bibinfo{pages}{133} (\bibinfo{year}{2022}), \eprint{2106.12825}.

\bibitem[{\citenamefont{Iancu and Mulian}(2021)}]{Iancu:2020mos}
\bibinfo{author}{\bibfnamefont{E.}~\bibnamefont{Iancu}} \bibnamefont{and}
  \bibinfo{author}{\bibfnamefont{Y.}~\bibnamefont{Mulian}},
  \bibinfo{journal}{JHEP} \textbf{\bibinfo{volume}{03}}, \bibinfo{pages}{005}
  (\bibinfo{year}{2021}), \eprint{2009.11930}.

\bibitem[{\citenamefont{Roy and Venugopalan}(2020)}]{roy:2019hwr}
\bibinfo{author}{\bibfnamefont{K.}~\bibnamefont{Roy}} \bibnamefont{and}
  \bibinfo{author}{\bibfnamefont{R.}~\bibnamefont{Venugopalan}},
  \bibinfo{journal}{Phys. Rev. D} \textbf{\bibinfo{volume}{101}},
  \bibinfo{pages}{034028} (\bibinfo{year}{2020}), \eprint{1911.04530}.

\bibitem[{\citenamefont{Hatta et~al.}(2022)\citenamefont{Hatta, Xiao, and
  Yuan}}]{hatta:2022lzj}
\bibinfo{author}{\bibfnamefont{Y.}~\bibnamefont{Hatta}},
  \bibinfo{author}{\bibfnamefont{B.-W.} \bibnamefont{Xiao}}, \bibnamefont{and}
  \bibinfo{author}{\bibfnamefont{F.}~\bibnamefont{Yuan}}
  (\bibinfo{year}{2022}), \eprint{2205.08060}.

\bibitem[{\citenamefont{Iancu et~al.}(2022)\citenamefont{Iancu, Mueller, and
  Triantafyllopoulos}}]{Iancu:2021rup}
\bibinfo{author}{\bibfnamefont{E.}~\bibnamefont{Iancu}},
  \bibinfo{author}{\bibfnamefont{A.~H.} \bibnamefont{Mueller}},
  \bibnamefont{and} \bibinfo{author}{\bibfnamefont{D.~N.}
  \bibnamefont{Triantafyllopoulos}}, \bibinfo{journal}{Phys. Rev. Lett.}
  \textbf{\bibinfo{volume}{128}}, \bibinfo{pages}{202001}
  (\bibinfo{year}{2022}), \eprint{2112.06353}.

\bibitem[{\citenamefont{Taels et~al.}(2022)\citenamefont{Taels, Altinoluk,
  Beuf, and Marquet}}]{Taels:2022tza}
\bibinfo{author}{\bibfnamefont{P.}~\bibnamefont{Taels}},
  \bibinfo{author}{\bibfnamefont{T.}~\bibnamefont{Altinoluk}},
  \bibinfo{author}{\bibfnamefont{G.}~\bibnamefont{Beuf}}, \bibnamefont{and}
  \bibinfo{author}{\bibfnamefont{C.}~\bibnamefont{Marquet}}
  (\bibinfo{year}{2022}), \eprint{2204.11650}.

\bibitem[{\citenamefont{Caucal et~al.}(2021)\citenamefont{Caucal, Salazar, and
  Venugopalan}}]{Caucal:2021ent}
\bibinfo{author}{\bibfnamefont{P.}~\bibnamefont{Caucal}},
  \bibinfo{author}{\bibfnamefont{F.}~\bibnamefont{Salazar}}, \bibnamefont{and}
  \bibinfo{author}{\bibfnamefont{R.}~\bibnamefont{Venugopalan}},
  \bibinfo{journal}{JHEP} \textbf{\bibinfo{volume}{11}}, \bibinfo{pages}{222}
  (\bibinfo{year}{2021}), \eprint{2108.06347}.

\bibitem[{\citenamefont{Bergabo and
  Jalilian-Marian}(2022{\natexlab{a}})}]{Bergabo:2021woe}
\bibinfo{author}{\bibfnamefont{F.}~\bibnamefont{Bergabo}} \bibnamefont{and}
  \bibinfo{author}{\bibfnamefont{J.}~\bibnamefont{Jalilian-Marian}},
  \bibinfo{journal}{Nucl. Phys. A} \textbf{\bibinfo{volume}{1018}},
  \bibinfo{pages}{122358} (\bibinfo{year}{2022}{\natexlab{a}}),
  \eprint{2108.10428}.

\bibitem[{\citenamefont{Bergabo and
  Jalilian-Marian}(2022{\natexlab{b}})}]{Bergabo:2022tcu}
\bibinfo{author}{\bibfnamefont{F.}~\bibnamefont{Bergabo}} \bibnamefont{and}
  \bibinfo{author}{\bibfnamefont{J.}~\bibnamefont{Jalilian-Marian}},
  \bibinfo{journal}{Phys. Rev. D} \textbf{\bibinfo{volume}{106}},
  \bibinfo{pages}{054035} (\bibinfo{year}{2022}{\natexlab{b}}),
  \eprint{2207.03606}.

\bibitem[{\citenamefont{Kovchegov
  et~al.}(2017{\natexlab{a}})\citenamefont{Kovchegov, Pitonyak, and
  Sievert}}]{Kovchegov:2017lsr}
\bibinfo{author}{\bibfnamefont{Y.~V.} \bibnamefont{Kovchegov}},
  \bibinfo{author}{\bibfnamefont{D.}~\bibnamefont{Pitonyak}}, \bibnamefont{and}
  \bibinfo{author}{\bibfnamefont{M.~D.} \bibnamefont{Sievert}},
  \bibinfo{journal}{JHEP} \textbf{\bibinfo{volume}{10}}, \bibinfo{pages}{198}
  (\bibinfo{year}{2017}{\natexlab{a}}), \eprint{1706.04236}.

\bibitem[{\citenamefont{Cougoulic and Kovchegov}(2019)}]{Cougoulic:2019aja}
\bibinfo{author}{\bibfnamefont{F.}~\bibnamefont{Cougoulic}} \bibnamefont{and}
  \bibinfo{author}{\bibfnamefont{Y.~V.} \bibnamefont{Kovchegov}},
  \bibinfo{journal}{Phys. Rev. D} \textbf{\bibinfo{volume}{100}},
  \bibinfo{pages}{114020} (\bibinfo{year}{2019}), \eprint{1910.04268}.

\bibitem[{\citenamefont{Kovchegov and Sievert}(2019)}]{Kovchegov:2018znm}
\bibinfo{author}{\bibfnamefont{Y.~V.} \bibnamefont{Kovchegov}}
  \bibnamefont{and} \bibinfo{author}{\bibfnamefont{M.~D.}
  \bibnamefont{Sievert}}, \bibinfo{journal}{Phys. Rev. D}
  \textbf{\bibinfo{volume}{99}}, \bibinfo{pages}{054032}
  (\bibinfo{year}{2019}), \eprint{1808.09010}.

\bibitem[{\citenamefont{Kovchegov
  et~al.}(2017{\natexlab{b}})\citenamefont{Kovchegov, Pitonyak, and
  Sievert}}]{Kovchegov:2017jxc}
\bibinfo{author}{\bibfnamefont{Y.~V.} \bibnamefont{Kovchegov}},
  \bibinfo{author}{\bibfnamefont{D.}~\bibnamefont{Pitonyak}}, \bibnamefont{and}
  \bibinfo{author}{\bibfnamefont{M.~D.} \bibnamefont{Sievert}},
  \bibinfo{journal}{Phys. Lett. B} \textbf{\bibinfo{volume}{772}},
  \bibinfo{pages}{136} (\bibinfo{year}{2017}{\natexlab{b}}),
  \eprint{1703.05809}.

\bibitem[{\citenamefont{Kovchegov
  et~al.}(2017{\natexlab{c}})\citenamefont{Kovchegov, Pitonyak, and
  Sievert}}]{Kovchegov:2016zex}
\bibinfo{author}{\bibfnamefont{Y.~V.} \bibnamefont{Kovchegov}},
  \bibinfo{author}{\bibfnamefont{D.}~\bibnamefont{Pitonyak}}, \bibnamefont{and}
  \bibinfo{author}{\bibfnamefont{M.~D.} \bibnamefont{Sievert}},
  \bibinfo{journal}{Phys. Rev. D} \textbf{\bibinfo{volume}{95}},
  \bibinfo{pages}{014033} (\bibinfo{year}{2017}{\natexlab{c}}),
  \eprint{1610.06197}.

\bibitem[{\citenamefont{Kovchegov
  et~al.}(2017{\natexlab{d}})\citenamefont{Kovchegov, Pitonyak, and
  Sievert}}]{Kovchegov:2016weo}
\bibinfo{author}{\bibfnamefont{Y.~V.} \bibnamefont{Kovchegov}},
  \bibinfo{author}{\bibfnamefont{D.}~\bibnamefont{Pitonyak}}, \bibnamefont{and}
  \bibinfo{author}{\bibfnamefont{M.~D.} \bibnamefont{Sievert}},
  \bibinfo{journal}{Phys. Rev. Lett.} \textbf{\bibinfo{volume}{118}},
  \bibinfo{pages}{052001} (\bibinfo{year}{2017}{\natexlab{d}}),
  \eprint{1610.06188}.

\bibitem[{\citenamefont{Kovchegov et~al.}(2016)\citenamefont{Kovchegov,
  Pitonyak, and Sievert}}]{Kovchegov:2015pbl}
\bibinfo{author}{\bibfnamefont{Y.~V.} \bibnamefont{Kovchegov}},
  \bibinfo{author}{\bibfnamefont{D.}~\bibnamefont{Pitonyak}}, \bibnamefont{and}
  \bibinfo{author}{\bibfnamefont{M.~D.} \bibnamefont{Sievert}},
  \bibinfo{journal}{JHEP} \textbf{\bibinfo{volume}{01}}, \bibinfo{pages}{072}
  (\bibinfo{year}{2016}), \bibinfo{note}{[Erratum: JHEP 10, 148 (2016)]},
  \eprint{1511.06737}.

\bibitem[{\citenamefont{Agostini
  et~al.}(2019{\natexlab{a}})\citenamefont{Agostini, Altinoluk, and
  Armesto}}]{Agostini:2019hkj}
\bibinfo{author}{\bibfnamefont{P.}~\bibnamefont{Agostini}},
  \bibinfo{author}{\bibfnamefont{T.}~\bibnamefont{Altinoluk}},
  \bibnamefont{and} \bibinfo{author}{\bibfnamefont{N.}~\bibnamefont{Armesto}},
  \bibinfo{journal}{Eur. Phys. J. C} \textbf{\bibinfo{volume}{79}},
  \bibinfo{pages}{790} (\bibinfo{year}{2019}{\natexlab{a}}),
  \eprint{1907.03668}.

\bibitem[{\citenamefont{Agostini
  et~al.}(2019{\natexlab{b}})\citenamefont{Agostini, Altinoluk, and
  Armesto}}]{Agostini:2019avp}
\bibinfo{author}{\bibfnamefont{P.}~\bibnamefont{Agostini}},
  \bibinfo{author}{\bibfnamefont{T.}~\bibnamefont{Altinoluk}},
  \bibnamefont{and} \bibinfo{author}{\bibfnamefont{N.}~\bibnamefont{Armesto}},
  \bibinfo{journal}{Eur. Phys. J. C} \textbf{\bibinfo{volume}{79}},
  \bibinfo{pages}{600} (\bibinfo{year}{2019}{\natexlab{b}}),
  \eprint{1902.04483}.

\bibitem[{\citenamefont{Altinoluk and Dumitru}(2016)}]{Altinoluk:2015xuy}
\bibinfo{author}{\bibfnamefont{T.}~\bibnamefont{Altinoluk}} \bibnamefont{and}
  \bibinfo{author}{\bibfnamefont{A.}~\bibnamefont{Dumitru}},
  \bibinfo{journal}{Phys. Rev. D} \textbf{\bibinfo{volume}{94}},
  \bibinfo{pages}{074032} (\bibinfo{year}{2016}), \eprint{1512.00279}.

\bibitem[{\citenamefont{Altinoluk
  et~al.}(2016{\natexlab{b}})\citenamefont{Altinoluk, Armesto, Beuf, and
  Moscoso}}]{Altinoluk:2015gia}
\bibinfo{author}{\bibfnamefont{T.}~\bibnamefont{Altinoluk}},
  \bibinfo{author}{\bibfnamefont{N.}~\bibnamefont{Armesto}},
  \bibinfo{author}{\bibfnamefont{G.}~\bibnamefont{Beuf}}, \bibnamefont{and}
  \bibinfo{author}{\bibfnamefont{A.}~\bibnamefont{Moscoso}},
  \bibinfo{journal}{JHEP} \textbf{\bibinfo{volume}{01}}, \bibinfo{pages}{114}
  (\bibinfo{year}{2016}{\natexlab{b}}), \eprint{1505.01400}.

\bibitem[{\citenamefont{Altinoluk et~al.}(2014)\citenamefont{Altinoluk,
  Armesto, Beuf, Mart\'\i{}nez, and Salgado}}]{Altinoluk:2014oxa}
\bibinfo{author}{\bibfnamefont{T.}~\bibnamefont{Altinoluk}},
  \bibinfo{author}{\bibfnamefont{N.}~\bibnamefont{Armesto}},
  \bibinfo{author}{\bibfnamefont{G.}~\bibnamefont{Beuf}},
  \bibinfo{author}{\bibfnamefont{M.}~\bibnamefont{Mart\'\i{}nez}},
  \bibnamefont{and} \bibinfo{author}{\bibfnamefont{C.~A.}
  \bibnamefont{Salgado}}, \bibinfo{journal}{JHEP}
  \textbf{\bibinfo{volume}{07}}, \bibinfo{pages}{068} (\bibinfo{year}{2014}),
  \eprint{1404.2219}.

\bibitem[{\citenamefont{Jalilian-Marian}(2021)}]{jalilian-marian:2021lhe}
\bibinfo{author}{\bibfnamefont{J.}~\bibnamefont{Jalilian-Marian}},
  \bibinfo{journal}{Nucl. Phys. A} \textbf{\bibinfo{volume}{1005}},
  \bibinfo{pages}{121943} (\bibinfo{year}{2021}).

\bibitem[{\citenamefont{Jalilian-Marian}(2020)}]{Jalilian-Marian:2019kaf}
\bibinfo{author}{\bibfnamefont{J.}~\bibnamefont{Jalilian-Marian}},
  \bibinfo{journal}{Phys. Rev. D} \textbf{\bibinfo{volume}{102}},
  \bibinfo{pages}{014008} (\bibinfo{year}{2020}), \eprint{1912.08878}.

\bibitem[{\citenamefont{Jalilian-Marian}(2019)}]{Jalilian-Marian:2018iui}
\bibinfo{author}{\bibfnamefont{J.}~\bibnamefont{Jalilian-Marian}},
  \bibinfo{journal}{Phys. Rev. D} \textbf{\bibinfo{volume}{99}},
  \bibinfo{pages}{014043} (\bibinfo{year}{2019}), \eprint{1809.04625}.

\bibitem[{\citenamefont{Jalilian-Marian}(2017)}]{Jalilian-Marian:2017ttv}
\bibinfo{author}{\bibfnamefont{J.}~\bibnamefont{Jalilian-Marian}},
  \bibinfo{journal}{Phys. Rev. D} \textbf{\bibinfo{volume}{96}},
  \bibinfo{pages}{074020} (\bibinfo{year}{2017}), \eprint{1708.07533}.

\bibitem[{\citenamefont{Hentschinski et~al.}(2018)\citenamefont{Hentschinski,
  Kusina, Kutak, and Serino}}]{Hentschinski:2017ayz}
\bibinfo{author}{\bibfnamefont{M.}~\bibnamefont{Hentschinski}},
  \bibinfo{author}{\bibfnamefont{A.}~\bibnamefont{Kusina}},
  \bibinfo{author}{\bibfnamefont{K.}~\bibnamefont{Kutak}}, \bibnamefont{and}
  \bibinfo{author}{\bibfnamefont{M.}~\bibnamefont{Serino}},
  \bibinfo{journal}{Eur. Phys. J. C} \textbf{\bibinfo{volume}{78}},
  \bibinfo{pages}{174} (\bibinfo{year}{2018}), \eprint{1711.04587}.

\bibitem[{\citenamefont{Hentschinski et~al.}(2016)\citenamefont{Hentschinski,
  Kusina, and Kutak}}]{Hentschinski:2016wya}
\bibinfo{author}{\bibfnamefont{M.}~\bibnamefont{Hentschinski}},
  \bibinfo{author}{\bibfnamefont{A.}~\bibnamefont{Kusina}}, \bibnamefont{and}
  \bibinfo{author}{\bibfnamefont{K.}~\bibnamefont{Kutak}},
  \bibinfo{journal}{Phys. Rev. D} \textbf{\bibinfo{volume}{94}},
  \bibinfo{pages}{114013} (\bibinfo{year}{2016}), \eprint{1607.01507}.

\bibitem[{\citenamefont{Gituliar et~al.}(2016)\citenamefont{Gituliar,
  Hentschinski, and Kutak}}]{Gituliar:2015agu}
\bibinfo{author}{\bibfnamefont{O.}~\bibnamefont{Gituliar}},
  \bibinfo{author}{\bibfnamefont{M.}~\bibnamefont{Hentschinski}},
  \bibnamefont{and} \bibinfo{author}{\bibfnamefont{K.}~\bibnamefont{Kutak}},
  \bibinfo{journal}{JHEP} \textbf{\bibinfo{volume}{01}}, \bibinfo{pages}{181}
  (\bibinfo{year}{2016}), \eprint{1511.08439}.

\bibitem[{\citenamefont{Balitsky and Tarasov}(2016)}]{Balitsky:2016dgz}
\bibinfo{author}{\bibfnamefont{I.}~\bibnamefont{Balitsky}} \bibnamefont{and}
  \bibinfo{author}{\bibfnamefont{A.}~\bibnamefont{Tarasov}},
  \bibinfo{journal}{JHEP} \textbf{\bibinfo{volume}{06}}, \bibinfo{pages}{164}
  (\bibinfo{year}{2016}), \eprint{1603.06548}.

\bibitem[{\citenamefont{Balitsky and Tarasov}(2015)}]{Balitsky:2015qba}
\bibinfo{author}{\bibfnamefont{I.}~\bibnamefont{Balitsky}} \bibnamefont{and}
  \bibinfo{author}{\bibfnamefont{A.}~\bibnamefont{Tarasov}},
  \bibinfo{journal}{JHEP} \textbf{\bibinfo{volume}{10}}, \bibinfo{pages}{017}
  (\bibinfo{year}{2015}), \eprint{1505.02151}.

\bibitem[{\citenamefont{Marquet et~al.}(2009)\citenamefont{Marquet, Xiao, and
  Yuan}}]{Marquet:2009ca}
\bibinfo{author}{\bibfnamefont{C.}~\bibnamefont{Marquet}},
  \bibinfo{author}{\bibfnamefont{B.-W.} \bibnamefont{Xiao}}, \bibnamefont{and}
  \bibinfo{author}{\bibfnamefont{F.}~\bibnamefont{Yuan}},
  \bibinfo{journal}{Phys. Lett. B} \textbf{\bibinfo{volume}{682}},
  \bibinfo{pages}{207} (\bibinfo{year}{2009}), \eprint{0906.1454}.

\bibitem[{\citenamefont{Kovchegov and Tuchin}(2002)}]{Kovchegov:2001sc}
\bibinfo{author}{\bibfnamefont{Y.~V.} \bibnamefont{Kovchegov}}
  \bibnamefont{and} \bibinfo{author}{\bibfnamefont{K.}~\bibnamefont{Tuchin}},
  \bibinfo{journal}{Phys. Rev. D} \textbf{\bibinfo{volume}{65}},
  \bibinfo{pages}{074026} (\bibinfo{year}{2002}), \eprint{hep-ph/0111362}.

\bibitem[{\citenamefont{Ayala et~al.}(2016)\citenamefont{Ayala, Hentschinski,
  Jalilian-Marian, and Tejeda-Yeomans}}]{ayala:2016lhd}
\bibinfo{author}{\bibfnamefont{A.}~\bibnamefont{Ayala}},
  \bibinfo{author}{\bibfnamefont{M.}~\bibnamefont{Hentschinski}},
  \bibinfo{author}{\bibfnamefont{J.}~\bibnamefont{Jalilian-Marian}},
  \bibnamefont{and} \bibinfo{author}{\bibfnamefont{M.~E.}
  \bibnamefont{Tejeda-Yeomans}}, \bibinfo{journal}{Phys. Lett. B}
  \textbf{\bibinfo{volume}{761}}, \bibinfo{pages}{229} (\bibinfo{year}{2016}),
  \eprint{1604.08526}.

\bibitem[{\citenamefont{Ayala et~al.}(2017)\citenamefont{Ayala, Hentschinski,
  Jalilian-Marian, and Tejeda-Yeomans}}]{ayala:2017rmh}
\bibinfo{author}{\bibfnamefont{A.}~\bibnamefont{Ayala}},
  \bibinfo{author}{\bibfnamefont{M.}~\bibnamefont{Hentschinski}},
  \bibinfo{author}{\bibfnamefont{J.}~\bibnamefont{Jalilian-Marian}},
  \bibnamefont{and} \bibinfo{author}{\bibfnamefont{M.~E.}
  \bibnamefont{Tejeda-Yeomans}}, \bibinfo{journal}{Nucl. Phys. B}
  \textbf{\bibinfo{volume}{920}}, \bibinfo{pages}{232} (\bibinfo{year}{2017}),
  \eprint{1701.07143}.

\bibitem[{\citenamefont{Balitsky}(1996)}]{Balitsky:1995ub}
\bibinfo{author}{\bibfnamefont{I.}~\bibnamefont{Balitsky}},
  \bibinfo{journal}{Nucl. Phys. B} \textbf{\bibinfo{volume}{463}},
  \bibinfo{pages}{99} (\bibinfo{year}{1996}), \eprint{hep-ph/9509348}.

\bibitem[{\citenamefont{Kovchegov}(2000)}]{Kovchegov:1999yj}
\bibinfo{author}{\bibfnamefont{Y.~V.} \bibnamefont{Kovchegov}},
  \bibinfo{journal}{Phys. Rev. D} \textbf{\bibinfo{volume}{61}},
  \bibinfo{pages}{074018} (\bibinfo{year}{2000}), \eprint{hep-ph/9905214}.

\bibitem[{\citenamefont{Jalilian-Marian
  et~al.}(1997{\natexlab{b}})\citenamefont{Jalilian-Marian, Kovner, Leonidov,
  and Weigert}}]{Jalilian-Marian:1997qno}
\bibinfo{author}{\bibfnamefont{J.}~\bibnamefont{Jalilian-Marian}},
  \bibinfo{author}{\bibfnamefont{A.}~\bibnamefont{Kovner}},
  \bibinfo{author}{\bibfnamefont{A.}~\bibnamefont{Leonidov}}, \bibnamefont{and}
  \bibinfo{author}{\bibfnamefont{H.}~\bibnamefont{Weigert}},
  \bibinfo{journal}{Nucl. Phys. B} \textbf{\bibinfo{volume}{504}},
  \bibinfo{pages}{415} (\bibinfo{year}{1997}{\natexlab{b}}),
  \eprint{hep-ph/9701284}.

\bibitem[{\citenamefont{Jalilian-Marian
  et~al.}(1998{\natexlab{a}})\citenamefont{Jalilian-Marian, Kovner, Leonidov,
  and Weigert}}]{Jalilian-Marian:1997jhx}
\bibinfo{author}{\bibfnamefont{J.}~\bibnamefont{Jalilian-Marian}},
  \bibinfo{author}{\bibfnamefont{A.}~\bibnamefont{Kovner}},
  \bibinfo{author}{\bibfnamefont{A.}~\bibnamefont{Leonidov}}, \bibnamefont{and}
  \bibinfo{author}{\bibfnamefont{H.}~\bibnamefont{Weigert}},
  \bibinfo{journal}{Phys. Rev. D} \textbf{\bibinfo{volume}{59}},
  \bibinfo{pages}{014014} (\bibinfo{year}{1998}{\natexlab{a}}),
  \eprint{hep-ph/9706377}.

\bibitem[{\citenamefont{Jalilian-Marian
  et~al.}(1998{\natexlab{b}})\citenamefont{Jalilian-Marian, Kovner, and
  Weigert}}]{Jalilian-Marian:1997ubg}
\bibinfo{author}{\bibfnamefont{J.}~\bibnamefont{Jalilian-Marian}},
  \bibinfo{author}{\bibfnamefont{A.}~\bibnamefont{Kovner}}, \bibnamefont{and}
  \bibinfo{author}{\bibfnamefont{H.}~\bibnamefont{Weigert}},
  \bibinfo{journal}{Phys. Rev. D} \textbf{\bibinfo{volume}{59}},
  \bibinfo{pages}{014015} (\bibinfo{year}{1998}{\natexlab{b}}),
  \eprint{hep-ph/9709432}.

\end{thebibliography}
\bibliographystyle{apsrev}

\end{document}